\documentclass{svjour3}                     
\smartqed  
\usepackage{graphicx}
\usepackage{natbib}
\usepackage[usenames, dvipsnames]{color}
\usepackage[colorlinks=true, linkcolor=blue, citecolor=blue, urlcolor=blue]{hyperref}
\usepackage{todonotes}
\usepackage{booktabs}
\usepackage{mathtools}
\usepackage{pifont}
\usepackage{mathrsfs}
\usepackage{bm}
\usepackage{amsmath}
\usepackage{amssymb}
\usepackage{csquotes}
\usepackage{upgreek}
\usepackage{graphicx}
\usepackage{caption}
\usepackage{subcaption}
\usepackage{todonotes}
\usepackage{euscript}
\usepackage{natbib}
\usepackage{lipsum}
\usepackage[margin=2.5cm]{geometry}
\usepackage{kantlipsum}

\usepackage{orcidlink}

\newcommand{\jfm}[1]{\textcolor{blue}{#1}}
\newcommand{\x}[1]{\textcolor{black}{#1}}
\newcommand{\vvg}[1]{\textcolor{black}{#1}}
\newcommand{\y}[1]{\textcolor{black}{#1}}
\newcommand{\rx}[1]{\textcolor{black}{#1}}

\definecolor{Saulo1}{RGB}{180,0,0}
\definecolor{Saulo11}{RGB}{230,0,0}
\definecolor{Saulo2}{RGB}{0,170,80}
\definecolor{Saulo3}{RGB}{0,230,100}
\definecolor{Saulo33}{RGB}{0,250,160}
\definecolor{Saulo4}{RGB}{0,160,240}
\definecolor{Saulo44}{RGB}{0,180,240}
\definecolor{Saulo5}{RGB}{120,240,250}
\definecolor{Saulo55}{RGB}{160,240,250}
\definecolor{Saulo6}{RGB}{200,240,250}
\definecolor{Saulo66}{RGB}{220,240,250}
\definecolor{Saulo7}{RGB}{190,190,190}
\definecolor{Saulo77}{RGB}{220,220,220}
\definecolor{Saulo777}{RGB}{240,240,240}

\begin{document}

\title{Global assessment of university research comprehensiveness
}

\author{Saulo Mendes\,\orcidlink{0000-0003-2395-781X}}

\institute{Saulo Mendes \at
              Group of Applied Physics and Institute for Environmental Sciences, University of Geneva, Boulevard Carl-Vogt 66, 1205 Geneva, Switzerland \\
              Tel.: +41-79-6264261 \\
              \email{saulo.dasilvamendes@unige.ch, saulomendesp@gmail.com}}

\date{\textit{Preprint in preparation for submission}}

\maketitle

\begin{abstract}

The demand for global university league tables has been high over the past two decades. However, significant criticism of their methodologies is accumulating without \rx{being addressed}. \x{One important bias regards the unequal distribution of research output and impact across different subjects, which in turn favors institutions that have its strongest performance in medical and physical sciences. I revisit global university league tables by normalizing each field as to create a uniform distribution of value.\rx{Then, the overall performance of an institution is interpreted as the probability of having a high score in any given academic field.} I focus on the similarity of institutions across ten criteria related to academic performance in eighty subjects of all fields of knowledge. The latter does not induce a \textit{zero-sum game}, removing one of the most prominent negative features of \y{established} league ta\y{bl}es.} The present assessment shows that the main difference between hundreds of leading global research universities is whether their coverage of all areas of human knowledge is comprehensive or specialized, \x{as their mean performance} \x{per subject is nearly indistinguishable}. I compare the results with the main league tables and found excellent agreement, suggesting that regardless of their methodologies, research-intensive institutions perform well in rankings if they are comprehensive. \rx{This comprehensiveness is ultimately dependent on institutional age, privileged funding allocation and regional academic culture. Consequently, when the size of an institution is taken out of the picture}, I found no correlation between comprehensiveness and quality, and no difference can be found in the \y{mean} quality of institutions regionally or globally. Furthermore, I find the reputation and prestige of \y{several} famous institutions to far exceed their performance within the present methodology, while \rx{numerous institutions with less reputation and visibility perform better than expected}.

\keywords{University \and Ranking \and Comprehensiveness \and Specialization \and Research}

\end{abstract}

\section{Introduction}

The modern university arose from the competition between the Jesuit, Oxbridge\footnote{\x{Term that regards collegiate universities that follow the management style for teaching and research as the universities of Oxford and Cambridge \citep{Tapper2002}.}} and German university frameworks \citep{clark2008,Menand2017}\x{, which} significantly differed from each other. While the Oxbridge model had no faculties and mostly focused on teaching for future fellows and administered by the head of the colleges, the German model focused on faculties of different academic subjects, had no colleges and was primarily focused on research \citep{Kerr2001,clark2008}. Confronted by the far more impactful German university model, even the famous Oxbridge and Paris academies \y{adopted} such system \citep{clark2008}. \x{During} the second-half of the nineteenth century, the American model combined the \x{Oxbridge and German} systems to create the \x{"}comprehensive\x{"} university \citep{Kerr2001,Zemliakova2018}. \x{Unlike the classical European universities, the American comprehensive university did not} reject technical subjects such as engineering, agriculture \x{or} commerce from its core curriculum. Furthermore, they combined the two systems in different ways: undergraduate studies at American universities followed the Oxbridge system while graduate studies followed the German research university model \citep{Veysey1965,crow2015}. It was therefore the American university that first combined nearly all subjects of human knowledge into a single campus, \x{initially} embodied in the prominent examples of \textit{Harvard}, \textit{Johns Hopkins} and \textit{Cornell} universities \citep{crow2015}. \x{Since then, \rx{these institutions outperformed smaller counterparts in drawing visibility due to hallmark discoveries in science, engineering, medicine or economics, so that}} it \rx{has become} natural \rx{to}  \x{deem the best} universities \rx{those with} strong research output and impact\x{. \rx{The association between university performance and research impact is very rooted,} to such an extent that global rankings are built to measure institutions against this model \citep{Marginson2007}}. Indeed, even teaching \x{in} graduate and undergraduate classrooms is ultimately related to an institution of research in the form of the "delivery of research-led teaching" \citep{taylor2006}. \x{Therefore,} it is only logical to measure university performance by research metrics \x{due to its} public availability and transparency. In fact, a comparative analysis among established university league tables show that although criteria vary significantly, research metrics appear in all of them and have the most prominent role among all indicators \citep{Buela-Casal2007}. This core research criteria make different league tables and methodologies thereof provide similar results \citep{Aguillo2010}.

\x{The evaluation of performance and prestige of universities can be traced back to} the 1798 \textit{Der Universit\"{a}ts-Bereiser} report to King Wilhelm II of Prussia \citep{Gedike2018}, regarded as one of the first forms of \x{regional} university ranking \citep{Menand2017}. \x{In the twentieth century, international institutional ranking did not exist in the present form, but the prestige of institutions were largely determined by awards, most promintently the Nobel prize \citep{Zuckerman1967,Inhaber1976,Zuckerman1977,Zuckerman1978}. On the other hand, the lack of a central education regulator led American institutions to compete among themselves for human and financial resources \citep{Hagstrom1971}, ultimately leading to a growing need for national rankings for both undergraduate and graduate studies, especially in the the post-war era \citep{Babcock1911,Cartter1966,webster1981,webster1986,Brankovic2021}. However, \x{due to the exponential growth of universities\footnote{\x{\citet{Buringh2009} estimates the existence of a dozen universities in medieval Europe. By the time of the industrial revolution it did not exceed a hundred universities, arguably a small number of institutions across many nations to prompt a need for rankings. In comparison, current worldwide estimates are placed around 30,000 universities \citep{Pavel2015}.}}, fast globalization of higher education and \rx{demand} for evaluation of cross-national tertiary educational systems \citep{Larsen2002,Merisotis2002,Schofer2005,Pavel2015},} the modern concept of global university ranking arose in 2003 with the \textit{\citet{Shanghai2022} Ranking} in an attempt by} China to compare its institutions to the leading American and European counterparts. \x{Especially because of the effect of university research on economic growth \citep{Anselin1997,Mokyr2011,Cantoni2014,Valero2019,Agasisti2020},} league tables have found a significant impact in geopolitics and policymaking \citep{Hazelkorn2008}, \x{and the competition it stems plays a major role in further enhancing the discourse on excellence of human development of nation states \citep{Brankovic2018}. Furthermore, rankings can influence the reshaping of national educational systems \citep{Marginson2007,Hazelkorn2015}. \x{Most prominently, the European tradition of \y{favoring} small specialized institutions have been recently challenged in view of the comprehensive university model, so that mergers of leading European instititutions have been carried out \citep{Docampo2015,Soler2019}.} \x{In addition, }university rankings are also revelant for the decision-making of} \rx{students} \x{regarding the} evaluation of the cost-benefit of choosing a particular institution \citep{Ehrenberg2002}. \x{Therefore,  assessments on university performance is undisputably essential for many economic and social spheres, both at the national and international level.}

\x{Despite the relevance of university performance assessments, significa\rx{n}t issues in the methodology of established league tables as well as the benchmarking of institutional quality has surfaced \y{\citep{Bowden2000,Clarke2002,Dill2005,Fauzi2020}}. For instance,} \citet{Ehrenberg2003} discussed the methodology of American college rankings, finding its main criteria to be based on reputation, selectivity of the institutions, retention and graduation rates, and even alumni donations. Although this rank still exists nationally, it is not surprising these factors are considered to be of the least importance when ranking global universities. \rx{Although} research-based rankings are more suitable \rx{than survey-oriented ones} for creating a global league table \citep{Taylor2007}, \x{global league tables relying on research-based metrics as provided by} the \textit{Shanghai Ranking} \x{also contain} issues and biases. \x{For instance, although recent versions of the \textit{Shanghai Ranking} have added awards and prizes in several other fields, their subject rankings do not include many social sciences as well as art and humanities subjects.}

\x{Building on the early criticisms of the \textit{Shanghai Ranking}, alternative league tables with new methodologies sprung in the same decade. Mostly as a mix of national rankings in the USA and research-led global ranking,} \textit{Quarelis Simons (QS)} and \textit{Times Higher Education Ranking (THE)}\footnote{\x{Available at \href{https://www.topuniversities.com/}{https://www.topuniversities.com} ; \href{https://www.timeshighereducation.com/}{https://www.timeshighereducation.com}}} \x{also relied on highly subjective parameters obtained from academic surveys, undergraduate employability, ratio of international students as well as class size.}
A major issue with employability indicators is that they are only applicable on a national or regional scale. Employers from South Africa can not compare its national institutions to those in New Zealand and so forth, as they do not compete together. Therefore, in an international ranking, employability will further skew scores to institutions that have a disproportionately favourable view from national employers as opposed to universities within countries where employability evaluation is uniform among the best institutions. In addition, employability favours schools who either specialize or score very high in engineering and decision sciences, such as law, economics or finance. Furthermore, it is well-known that non-academic employability of pure and theoretical physical sciences, humanities, social sciences are very low \citep{Garrouste2014}. As already argued in \citet{Ehrenberg2003} and \citet{Clarke2002}, these indicators are very problematic, as they are either measuring performance of previous decades or rooted in privilege and reputation that may not actually reflect the current state of an institution. Naturally, these survey-based indicators have found to lead to data fabrication, markedly exhibited in the recent scandals involving \textit{Temple Business School}\footnote{ See \textit{"Former Temple Business-School Dean Gets Prison Term in Rankings Scandal" at} \href{https://www.wsj.com/articles/former-temple-business-school-dean-gets-prison-term-in-rankings-scandal-11647053211}{Wall Street Journal}. } and \textit{Columbia University}\footnote{ \x{ See the article \textit{"Columbia is the latest university caught in a rankings scandal"} at \href{ https://www.economist.com/united-states/columbia-is-the-latest-university-caught-in-a-rankings-scandal/21808445 }{ The Economist }.}}, in addition to being found to have significant conflicts of interest \citep{Chirikov2021}.

\x{Yet a third group of global university rankings arose:\footnote{\x{Among those not listed, webometrics is one the earliest. However, it is mostly based on internet transparency, not research or similar metrics. The league tables are found at \href{http://nturanking.csti.tw/}{NTU Ranking}, \href{https://www.usnews.com/education/best-global-universities/rankings}{US News \& Report Global University Rankings}, \href{https://www.leidenranking.com/}{Leiden Ranking}, \href{https://www.scimagoir.com/rankings.php?sector=Higher\%20educ.}{SCImago Ranking}. }}} \x{the pure\y{ly} bibliometric rankings such as the \textit{NTU Ranking}, \textit{Leiden Ranking}, \textit{US News \& Report Global University Rankings} among others. The main feature of these rankings is the abhorrence towards surveys and reputation of non-research metrics, and almost all indicators are related to bibliometrics. Indeed, \citet{Chen2012} showed that \textit{Shanghai Ranking} correlated well with bibliometric-based rankings, and survey-based ones showed the least overlap. Not only the overlap is less significant when comparing research-based with survey and reputation-based rankings, the latter tend to be prone to lack of transparency, gaming of data or total fabrication that can not be assessed by independent examiners. On the other hand, research-based metrics tend to be transparent and reproducible \citep{Docampo2013}. Nevertheless, pure research basis have also shown some setbacks:} it is well-known that established university league tables favors universities who have a very high research output in natural \x{and} medical sciences, \x{as these fields claim the journals with highest impact factor and highest number of published articles. \y{Indeed}, a \textit{Matthew effect} \citep{Biglu2008} was found on the distribution of research output across different fields, \y{with} fields \y{of} highest impact claim\y{ing} the highest growth in impact factor \citep{Althouse2009}. Hence, there is little hope that social sciences, humanities and engineering will ever reach the sheer number of citations and published articles of natural and medical sciences \citep{Hamilton1990}.} \y{In fact,} institutions can have their \y{rank} uplifted if specific scientific fields have a higher weight measure in the computation of the ranking, either by active methodological choice or by unconscious bias towards fields that provide higher metrics. \y{In this context, i}nstitutions whose strongest departments are in the social sciences and humanities \y{are severely underestimated. Conversely}, having \y{strong} medical \y{and} science departments \y{is} enough to \y{maintain or attract} prestige. Therefore, a fair measurement of university rankings \y{require} a proper \x{subject} normalization.
\begin{figure}
\centering
    \includegraphics[scale=0.17]{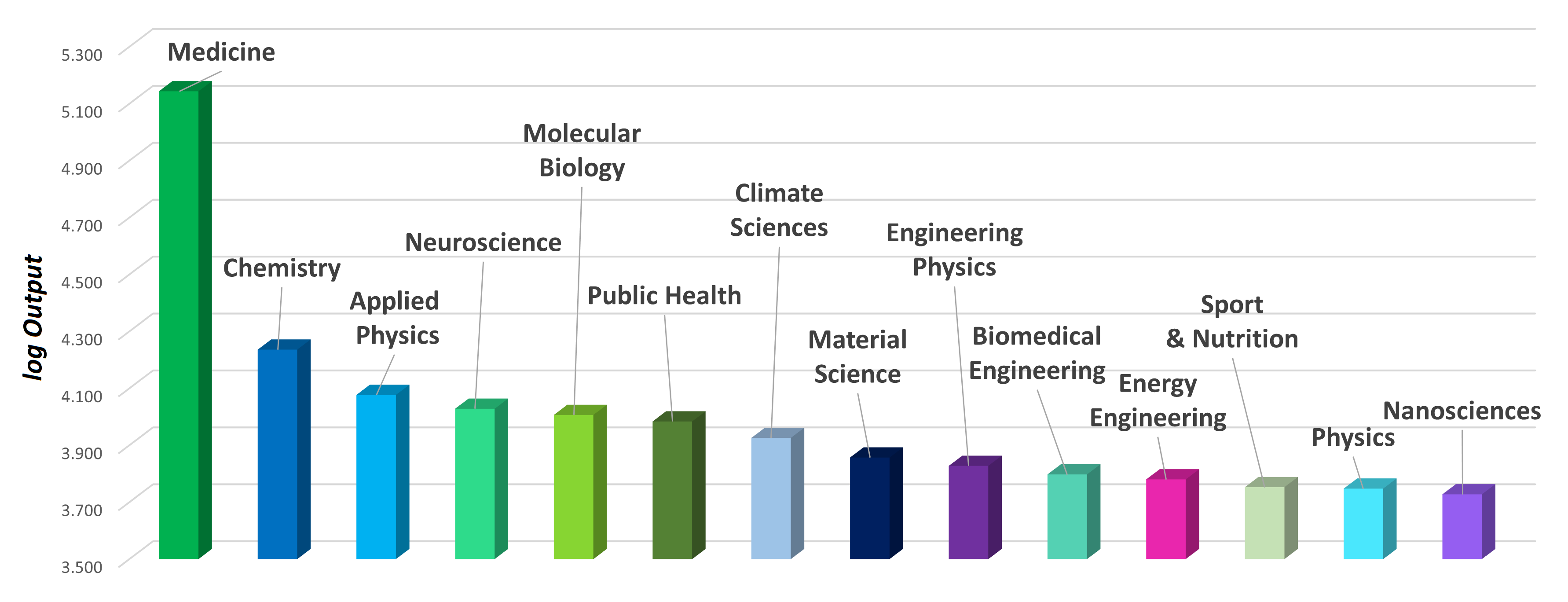}
\caption{\y{List of academic} subjects (see \jfm{Table} \ref{tab:subjects}) \y{ranked} by research output (logarithm of the number of indexed documents in \textit{web of Science} in the period 2011-2020) of the\y{ir leading} institution\y{s}. The leader in the subject of medicine (\textit{Harvard University}) published 139,000 documents in this period, while the leader in physics (\textit{Universit\'{e} de Paris-Saclay}) published 6,000 documents and the leader in chemistry (\textit{University of the Chinese Academy of Sciences}) published over 17,000 documents.}
\label{fig:TopSubject}
\end{figure}
\begin{table*}
\centering
\begin{tabular}{l|rrrrr|l}  
\toprule
\emph{Institution} &  \emph{QS Rank 2020}  &  \emph{\#Top50 QS}   &  \emph{CWUR 2017} &  \emph{\#Top10 CWUR} & \bf{\emph{QS+CWUR}} &  \emph{Country}
\\
\midrule
 Harvard                  & \bf{3}   & 35 & \bf{1}  & 112 & \bf{\textit{147}} & United States  \\
 UC Berkeley              & \bf{28}  & 38 & \bf{7}  & 50  & \bf{\textit{ 88}} & United States  \\
 Stanford                 & \bf{2}   & 38 & \bf{2}  & 48  & \bf{\textit{ 86}} & United States  \\
 MIT                      & \bf{1}   & 29 & \bf{3}  & 41  & \bf{\textit{ 70}} & United States  \\
 Princeton                & \bf{13}  & 26 & \bf{9}  & 9   & \bf{\textit{ 35}} & United States  \\
 Caltech                  & \bf{5}   & 12 & \bf{11} & 8   & \bf{\textit{ 20}} & United States  \\
 \midrule           
 Oxford                   & \bf{4}   & 38 & \bf{5}  & 47  & \bf{\textit{ 85}} & United Kingdom \\
 Cambridge                & \bf{7}   & 39 & \bf{4}  & 38  & \bf{\textit{ 87}} & United Kingdom \\
 UCL                      & \bf{8}   & 33 & \bf{31} & 37  & \bf{\textit{ 70}} & United Kingdom \\
 \midrule            
 Sorbonne                 & \bf{77}  & 9  & \bf{56} & 17  & \bf{\textit{ 26}} & France         \\
 \'{E}cole Poly.          & \bf{60}  & 3  & \bf{35} & 0   & \bf{\textit{  3}} & France         \\
 \midrule 
\bottomrule
\end{tabular}
\caption{Comparison between world ranking and the number of subjects in which universities are world leaders from \textit{QS Ranking 2020} and \textit{CWUR Ranking 2017}. The former measured university performance in all areas of human knowledge subdivided into 46 subjects, while the latter had 227 subjects due to the inclusion of subareas of all natural and medical sciences.}
\label{tab:group0}
\end{table*}

\vvg{\y{Despite} the long list of issues present in most league tables, \y{a} concrete comparison between a large number of universities can be made based on research impact and comprehensiveness. However,} as \vvg{shown} in \jfm{Figure} \ref{fig:TopSubject}, some subjects have a much higher research output \vvg{(articles, reviews, etc.)} than others. Moreover, they also vary wildly in citation metrics and journal impact factor \citep{Althouse2009}. Overall, bibliometric practices \vvg{strongly vary among different subjects} \citep{Moed1985} and makes \vvg{any global evaluation based on weighted metrics} \y{biased} and unrealistic. \vvg{Albeit} articles have steadily increased their number of references over the last decades, \vvg{their} grow\vvg{th} \vvg{in} STEM fields \vvg{is bigger} than \vvg{in} social sciences and humanities \citep{Biglu2008,Dai2021} and \vvg{the} variability across fields further widens the unevenness of their bibliometrics \citep{Seglen1992}. Hence, league tables based on total \vvg{weighted} research \vvg{are inevitably favorable to} \y{the leading} institutions in medi\vvg{cal}, chemi\vvg{cal} and physi\vvg{cal sciences}.

\jfm{Table} \ref{tab:group0} displays a much less discussed feature of university league tables: the \textit{non-extensive} character that creates an enormous disparity between excellence in academic subjects and the overall excellence. Although \textit{UC Berkeley} is clearly a peer of \textit{Harvard, Stanford} and \textit{MIT} by the measure of the number of subjects they excel in, the \textit{QS} overall ranking seems to contradict its own subject evaluation  and puts \textit{UC Berkeley} far below its peers in the overall analysis. Furthermore, the \textit{Sorbonne \y{Universit\'{e}}} seems to be at the same overall rank as the \textit{\'{E}cole Polytechnique} even though the former excels in dozens of subjects and the latter in only a few. Strikingly, the two overall rankings suggest that \textit{Caltech} is far ahead of \textit{Sorbonne}, while the latter actually has similar subject performance of \textit{Caltech}. One can not help but perceive this phenomenon as paradoxical to the claim of academic excellence. Other \y{league} tables such as the \textit{Shanghai Ranking} feature the same problem. The methodology \y{of} established rankings create this paradox by selecting \textit{intensive} criteria that favours a few subjects, either present in bibliometric or reputation criteria. To remove this conflicting issue, in this work I analyze the comprehensiveness and evaluate the overall research impact of an institution as \textit{extensive} (henceforth in the paper I take it to mean an additive property), i.e. the ideal \y{(or maximal)} research-intensive university is axiomatically the one which excels in all subjects.

\x{In addressing all these points, t}he present study focuses and creates a method to rank global institutions \x{limited to research-based indicators extracted from publicly available data of \textit{web of Science}.} \x{In the spirit of \citet{Kosztyan2019}, \y{I replace unidimensional} ranking\y{s by well-posed bideminesional ones:} a collection of \y{ranked} leagues with\y{in the otherwise full unidimensional} ranking. \y{However,} the present model define these groups by the similarity of these institutions across twelve indicators derived from each of the eighty subjects. In doing so, I address \y{the} well-known pathology of \textit{zero-sum games} \citep{Lee2020}} and avoid \x{the practice of} gaming \x{or tweaking} of \x{indicators}. \x{This new method of normalized and extensive analysis where the total score is a reflection of how many subjects an institution excels, shows that institutions are overvalued due to either their financial privileges, reputation, prestigious awards or disproportionate high performance in subjects with the highest bibliometric impact. Remarkably, the mean ranking among \y{five leading} league tables is shown to agree well with the present model, which confirms the tangible existence of the stratification of academic excellence.}

\section{Extensive World Rankings: Subject Comprehensiveness}\label{sec:methods}

\begin{figure}
\centering
    \includegraphics[scale=0.24]{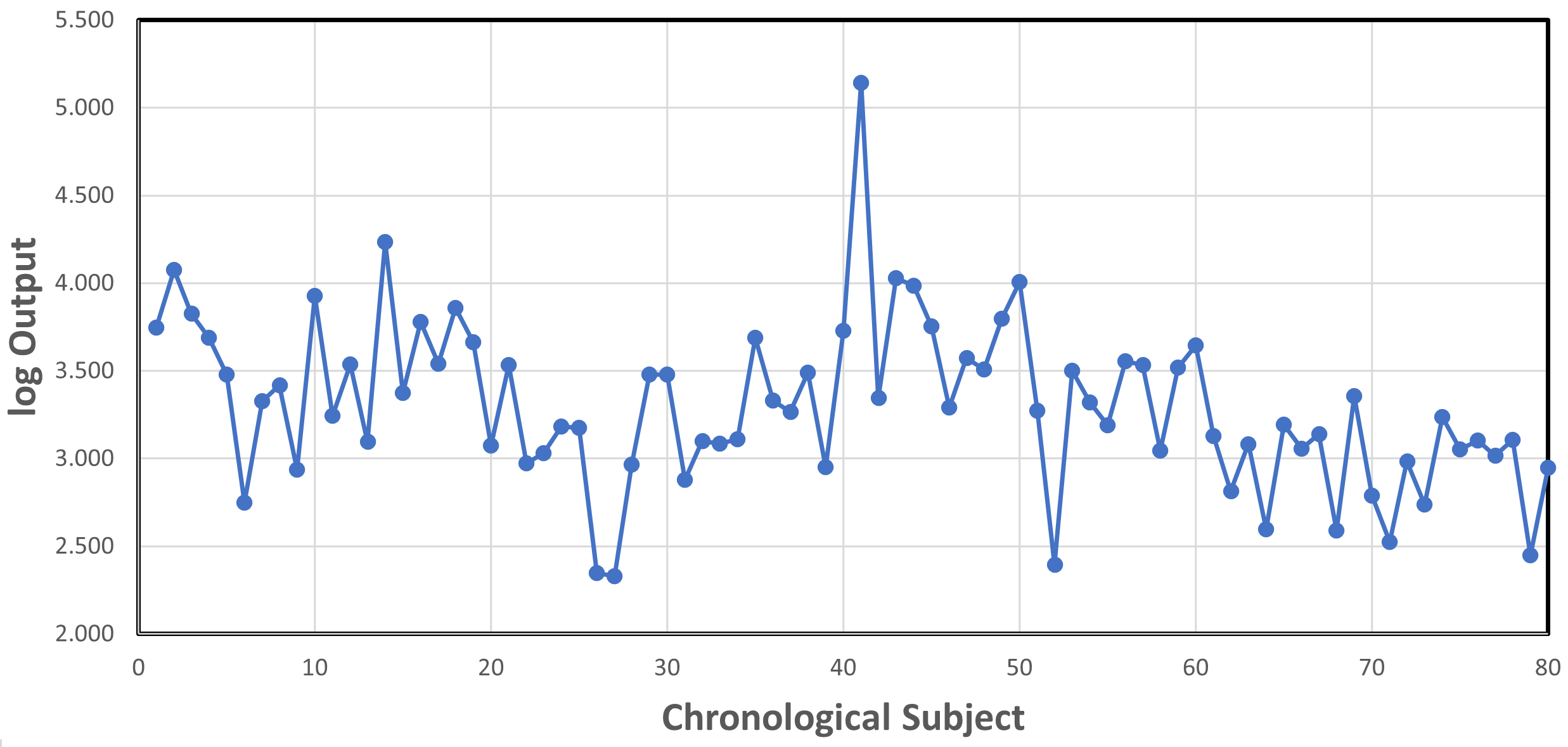}
\caption{Distribution of subjects according to the highest research output in $log_{10}$ scale of the best institution for each subject, as ordered in \jfm{Table} \ref{tab:subjects}.}
\label{fig:SubjectSeries}
\end{figure}
\begin{table}
\begin{tabular}{lllll} 
\toprule
 1 \, Physics           &  2 \, Applied Physics     &   3 \, Eng. physics           &   4 \, Astronomy               &  5 \, Mathematics       \\ 
 6 \, Statistics          &  7 \, Geophysics        &   8 \, Meteorology            &   9 \, Phys. Oceanography      & 10 Climate Science      \\ 
11 Geology             & 12 Geochemistry      &  13 Chem. Oceanography     &  14 Chemistry               & 15 Chemical Eng.     \\ 
16 Energy Eng.         & 17 Petroleum Eng.    &  18 Material Science          &  19  Metallurgical Eng.     & 20 Mineral Eng.      \\ 
\midrule
21 Civil Eng.          & 22 Geological Eng.   &  23 Transport Eng.         &  24  Environmental Eng.     & 25 Hydraulic Eng.    \\ 
26 Coastal Eng.        & 27 Architecture      &  28 Urban Planning         &  29 Geodesy                 & 30 Mechanical Eng.   \\ 
31 Naval Eng.          & 32 Aerospace Eng.    &  33 Industrial Eng.        &  34 Acoustic Eng.           & 35 Electrical Eng.   \\ 
36 Computer Sci.       & 37 Computer Eng.     &  38 Telecom. Eng. &  39  Mechatronical Eng.     & 40 Nanoscience       \\
\midrule
41 Medicine            & 42 Nursing           &  43 Neurosci. \& Psych.           &  44  Public Health          & 45 Sport \& Nutrition \\ 
46 Phys. Therapy       & 47 Dentistry         &  48 Pharmacy               &  49  Biomedical Eng.        & 50 Molecular Biology   \\ 
51 Biochemistry        & 52 Biophysics        &  53 Biology                &  54 Ecology                 & 55 Biol. Geosciences \\ 
56 Agriculture   & 57 Food Eng.         &  58 Biotechnology          &  59 Vet. Medicine           & 60 Plant Science        \\ \midrule
61 Economics           & 62 Finance           &  63 Business               &  64 Public Policy           & 65 Law               \\ 
66 Politics            & 67 Sociology         &  68 Anthropology           &  69 Education               & 70 Journalism        \\ 
71 Media studies       & 72 Management        &  73 Archaeology            &  74  History                & 75 Geography         \\ 
76 Philosophy          & 77 Theology          &  78 Languages              &  79 Music                   & 80 Arts              \\ 
\bottomrule
\end{tabular}
\caption{The partition of eighty \y{academic} subjects for the evaluation of university performance between the fields of \rx{pure and applied} physical, earth \& chemical sciences (1-20), engineering sciences (21-40), \rx{biological} \& life sciences (41-60) and social sciences \& humanities (61-80).}
\label{tab:subjects}
\end{table}
\vvg{In view of the \y{several} caveats \y{emerging in the process of classifying} universit\y{}ies, I \y{propose to} ammend these issues by measuring the groupiness or similarity of institutions. \y{I} separate institutions by groups of similarity rather than a continuous ranking. Below I describe the step-by-step methodology:}

\begin{displayquote}

\vvg{
\textbullet \, Delineate subareas for each of the eighty subjects according to \textit{web of Science} classification over the period between 2011-2020.
}\vspace{+0.1cm}

\vvg{
\textbullet \, Separate a measure of research impact \y{applicable to \rx{each} subject. T}he \textit{Leiden Rank} parameter $PP_{10\%}$, henceforth denoted by $\mathcal{L}$ and \y{normalized to account for subject differences}, measur\y{es} the percentage of an institution output \y{at} the top 10\% \y{of the} most cited papers \y{and is the most suitable measure of research impact due to its weak dependence on time, as opposed to average citations \citep{Sangwal2011,Finardi2014} or the \citeauthor{Hirsch2005}'s h-index\footnote{\y{Note that the time-dependence of the h-index was already discussed in \citet{Hirsch2005}, where the author proposed to normalize the h-index by one's net academic age.}} \citep{Egghe2007,Mannella2013}}. The fractional counting \y{is applied to} remove the effect of large collaborations in medical and physical sciences. This parameter superseeds \y{average} citations \y{as it weakens the citation disparity among subjects}.
}\vspace{+0.1cm}

\vvg{
\textbullet \, For multi-campus universities I only measure the performance of the main campus. This is due the fact that leading US institutions have \y{several} campuses that are not computed altogether. \y{For instance, were I to compute all campuses}, the \textit{University of California} output would far exceed the peers of its flagship campus (\textit{UC Berkeley}) by an order of magnitude. Therefore, I define the parameter $\rho$ as the percentage of the research output carried out by the main campus of an institution, with $\rho \approx 1$ in the vast majority of cases.
}\vspace{+0.1cm}

\vvg{
\textbullet \, For each $i$-th subject with $1 \leqslant i \leqslant 80$, I compute the total research output between 2011-2020 as the sum of all citeable documents (articles, reviews, conference papers, etc.). I henceforth denote this number by $\mathcal{O}_{i}$. The arbitrary chronological $i$-order is given in \jfm{Table} \ref{tab:subjects}. Each subject can contain at most 200 ranked institutions, thus listing only the world \y{1\%} leading universities in a given subject.
}\vspace{+0.1cm}

\vvg{
\textbullet \, For each $i$-th subject I compute the maximum output and denote it with $\mathbb{E}[\mathcal{O}_{i}]$.
}\vspace{+0.1cm}

\vvg{
\textbullet \, The $i$-th subject has a score $0 \leqslant \Lambda_{ik} \leqslant 1000$ for the $k$-th institution, and is computed with normalization $A_{i}$:}
\begin{equation}
\vvg{ \Lambda_{ik} := A_{i} \sqrt{\rho \mathcal{L} \cdot \frac{\mathcal{O}_{ik}}{\mathbb{E}[\mathcal{O}_{ik}]}}
} \quad \y{ \therefore \quad   \Lambda_{ik} \left( \mathbb{E}[\mathcal{O}_{i}]  \right) \equiv 1000 \quad .   }
\end{equation}
\vspace{+0.1cm}

\vvg{
\textbullet \, The major variable $0 \leqslant X \leqslant 80,000$ measures the total score of all covered subejcts (appearing among the top 200 institutions) by an institution removing the output bias of each subject:}
\begin{equation}
\vvg{ X := \sum_{i=1}^{80} A_{i} \sqrt{\rho \mathcal{L} \cdot \frac{\mathcal{O}_{i}}{\mathbb{E}[\mathcal{O}_{i}]}}
} \quad .
\end{equation}
\vspace{+0.1cm}

\vvg{
\textbullet \, The major variable $0 \leqslant Y \leqslant 1000$ measures the total score of all covered subejcts by an institution \y{regardless of its subject rank}, though keeping the output bias of each subject:}
\begin{equation}
\vvg{ Y :=  A \sqrt{\rho \mathcal{L} \cdot \left( \sum_{i=1}^{80} \mathcal{O}_{i} \right)^{1/2} }
} \quad .
\end{equation}
\vvg{The root of the output appears because the weighted sum of all subject outputs have only one normalization constant A and the exponential decay in output is much faster than for X.}
\vspace{+0.1cm}

\vvg{
\textbullet \, The variable $0 \leqslant \overline{\Lambda_{X}} := X/\mathcal{N}_{200} \leqslant 1000$ computes the arithmetic mean among all subjects, where $\mathcal{N}_{j}$ measur\y{es} the number of subjects $i$ in which a given institution appear among the top $j$ universities. This parameter is the \y{closest measure of} institutional research \y{"quality"}.
}\vspace{+0.1cm}

\vvg{
\textbullet \, The variable $0 \leqslant \overline{\Lambda_{Y}} :=(X+80Y)/(\mathcal{N}_{200}+80) \leqslant 1000$ computes the weighted mean among all subjects \y{regardless of its subject rank}.
}\vspace{+0.1cm}

\vvg{
\textbullet \, The variable $0 \leqslant \overline{\Lambda_{\mathcal{N}}} := B \cdot \overline{\Lambda_{Y}} \tanh{[ (1 + \mathcal{N}_{200} ) / 30  ]}\leqslant 1000$ adjusts the mean $\overline{\Lambda_{Y}}$ to the number $i$ of subjects they excel at, \y{where} $B \approx 1$ is a normalization constant. It differentiates institutions posessing the same mean $\overline{\Lambda_{Y}}$ but that are indeed far apart in the measure of comprehensiveness.
}\vspace{+0.1cm}

\vvg{
\textbullet \, The variable $0 \leqslant \mathcal{C} := 1000 (\mathcal{N}_{200}+\mathcal{N}_{100}+\mathcal{N}_{30}+\mathcal{N}_{1}+1) /( 3\mathcal{N}_{200} + 81 )  \leqslant 1000$ measures the degree of comprehensiveness concurrent with research impact of an institution.
}\vspace{+0.1cm}

\vvg{
\textbullet \, The auxiliary variable $\varphi$ measures the mean score per research faculty of an institution:
\begin{equation}
\varphi := \frac{ 80 \overline{\Lambda_{Y}} }{ 2000 \tanh{(M_{F}/2500)} } \quad ,
\end{equation}
where $M_{F}$ is the total number of permanent faculty. Note that \y{if} $M_{F} \ll 10,000$ the effective number of research-intensive faculty converges to $(4/5)M_{F}$. \y{This asymptotic limit} account\y{s} for the known fact that teaching stream positions (i.e. not involved in research \y{and restricted to teaching duties}) in research-intensive institutions \y{do not exceed} 20\% of \y{the total full-time} faculty \y{count}\footnote{\y{Excluding medical schools, research-intensive institutions in the United States tend to restrict teaching-only faculty positions in the non-tenure track. Table 3.1 of \citet{Ehrenberg2005} shows that by 1999 about 15\% of full-time faculty were of non-tenure track. However, this number has slowly increased in the years since \citep{Curtis2014}. As a remark, the research-oriented faculty does not consist of research-only duties. On the contrary, their teaching and administrative load is typically higher than of research \citep{Schuster2006}}.}. \y{In order to focus on research metrics while not upholding any bias towards socialized higher educational systems (which have an inflated number of teaching-only faculty),} the above formula \y{removes} teaching stream instructors \y{from the total number of full-time faculty}. 
}\vspace{+0.1cm}

\begin{table*}
  \centering
\begin{tabular}{r|rrrrrrrrrrrrrrr}   
\toprule
\emph{Group/Color} &
 \emph{$ X $} &  \emph{$X^{\ast}$} &  \emph{$Y$}  &  \emph{$Y^{\ast}$} &  \emph{$\overline{\Lambda_{X}}$}  &  \emph{$\overline{\Lambda_{Y}}$}  &   \emph{$\overline{\Lambda_{\mathcal{N}}}$}  &  \emph{$\overline{\Lambda^{\ast}_{\mathcal{N}}}$} & \emph{$\mathcal{C}$}   &  \emph{$\mathcal{C}^{\ast}$}
\\
\midrule
\colorbox{Saulo1}{\qquad 1 \qquad} & \vvg{40,000} & \vvg{40,000} & \vvg{40,000} & \vvg{40,000} & \vvg{600} & \vvg{600} & \vvg{550} & \vvg{550} & \vvg{550} & \vvg{550}  \\
\colorbox{Saulo2}{\qquad 2 \qquad} & \vvg{30,000} & \vvg{30,000} & \vvg{30,000} & \vvg{30,000} & \vvg{500} & \vvg{500} & \vvg{450} & \vvg{450} & \vvg{450} & \vvg{450}  \\
\colorbox{Saulo3}{\qquad 3 \qquad} & \vvg{20,000} & \vvg{20,000} & \vvg{20,000} & \vvg{20,000} & \vvg{400} & \vvg{400} & \vvg{350} & \vvg{350} & \vvg{350} & \vvg{350}  \\
\colorbox{Saulo4}{\qquad 4 \qquad} & \vvg{10,000} & \vvg{10,000} & \vvg{10,000} & \vvg{10,000} & \vvg{350} & \vvg{350} & \vvg{250} & \vvg{250} & \vvg{250} & \vvg{250}  \\
\colorbox{Saulo5}{\qquad 5 \qquad} & \vvg{6,000}  & \vvg{6,000}  & \vvg{6,000}  & \vvg{6,000}  & \vvg{300} & \vvg{300} & \vvg{150} & \vvg{150} & \vvg{150} & \vvg{150}  \\
\colorbox{Saulo6}{\qquad 6 \qquad} & \vvg{3,000}  & \vvg{3,000}  & \vvg{3,000}  & \vvg{3,000}  & \vvg{250} & \vvg{250} & \vvg{80}  & \vvg{80}  & \vvg{80}  & \vvg{80}   \\
\colorbox{Saulo7}{\qquad 7 \qquad} & \vvg{1,000}  & \vvg{1,000}  & \vvg{1,000}  & \vvg{1,000}  & \vvg{200} & \vvg{200} & \vvg{20}  & \vvg{20}  & \vvg{20}  & \vvg{20}   \\
\bottomrule
\end{tabular}
\caption{\vvg{Thresholds for the set of variables that defines mean groupiness and peer similarity. The remaining group 8 does not require a minimum threshold. Integer groups are assigned with a unique color to portray \y{group} stratification as in \jfm{Figure} \ref{fig:OGamma}.}}
\label{tab:groupiness}
\end{table*}

\vvg{
\textbullet \, The auxiliary variable $\varphi$ transforms the size-dependent variables $\mathcal{V}= (X, Y, \overline{\Lambda_{\mathcal{N}}}, \mathcal{C})$ into size-independent ones $\mathcal{V}^{\ast}= (X^{\ast}, Y^{\ast}, \overline{\Lambda^{\ast}_{\mathcal{N}}}, \mathcal{C}^{\ast})$ through the change of variables $\mathcal{V}^{\ast} = \mathcal{V} \cdot \sqrt{\varphi /40}$. The square root and normalization $1/40$ appear to maintain the size-independent variables in the same range [0,1000]. Note that for an institution obtaining $\Lambda_{1}=\Lambda_{2}= \dots = \Lambda_{i} = \dots =\Lambda_{80}=1000$, currently not existing, would require a very large $M_{F}$ and therefore $\lim_{M_{F} \rightarrow \infty } \varphi (\Lambda_{i} = 1000 \, , \, \forall i) = 40$. 
}\vspace{+0.1cm}

\vvg{
\textbullet \, I combine \y{five} faculty size-independent variables $\mathcal{V}^{\ast}= (X^{\ast}, Y^{\ast}, \overline{\Lambda_{X}}, \overline{\Lambda^{\ast}_{\mathcal{N}}}, \mathcal{C}^{\ast})$ and \y{five} faculty size-dependent variables $\mathcal{V}= (X, Y, \overline{\Lambda_{Y}}, \overline{\Lambda_{\mathcal{N}}}, \mathcal{C})$ to create the set $(\mathcal{V},\mathcal{V}^{\ast})$ defining groupiness (\textit{peer similarity}). For each variable, I define a minimum threshold \y{for an institution} to belong to a group $G_{1 + \frac{p}{2}}$. Fractional groups exist as an ammend to institutions that do not obey the minimum criteria to belong to a group $G_{p+1}$ but that are not peer similar to any institution in the next integer group $G_{p+2}$. To make this groupiness analysis robust, I oscillate the threshold by $\pm 5\%$ and $\pm 10\%$. Therefore, each institution will be assigned a groupiness number $G_{1 + \frac{p}{2}}$ for each variable of the set $(\mathcal{V}_{\pm},\mathcal{V}^{\ast}_{\pm})$. The thresholds without oscillation\y{s} is shown in \jfm{Table} \ref{tab:groupiness}.
}\vspace{+0.1cm}

\vvg{
\textbullet \, I compute the mean groupiness $\langle G \rangle$ among all fifty values of the set $(\mathcal{V}_{\pm},\mathcal{V}^{\ast}_{\pm})$. The original five major variables were expanded for robustness, enlarged twofold to account for faculty size and then fivefold for threshold flexibility. Let $(n,p) \in \mathbb{N}$ with $n > 0$ \y{with $p$ odd}, an institution $U$ belong\y{s} to a group $G_{n+\frac{p}{2}}$ \y{provided that}: 
\begin{equation}
U \in G_{n+\frac{p}{2}} \iff \langle G \rangle \leqslant \left( n +\frac{p}{2} + 1 \right)  \quad \textrm{and} \quad \frac{\sum_{n,p}(G_{n+\frac{p}{2}}+G_{n+\frac{p}{2}-1})}{\sum_{n,p}G_{\forall n,p}} \geqslant 0.5 \,\, .
\label{eq:G1}
\end{equation} 
On the other hand, any institution $U$ will neither belong to group $G_{n+\frac{p}{2}}$ nor $G_{n+\frac{p}{2}+1}$ as long as,
\begin{equation}
U \in G_{n+\frac{(p+1)}{2}} \iff \langle G \rangle \leqslant \left( n +\frac{p}{2} + \frac{3}{2} \right) \quad \textrm{and} \quad \frac{\sum_{n,p}(G_{n+\frac{p}{2}}+G_{n+\frac{p}{2}+1})}{\sum_{n,p}G_{\forall n,p}} \geqslant 0.95 \,\, .
\label{eq:G2}
\end{equation} 
}\vspace{+0.1cm}

\begin{figure}
\centering
    \includegraphics[height=1.49cm,width=16.0cm]{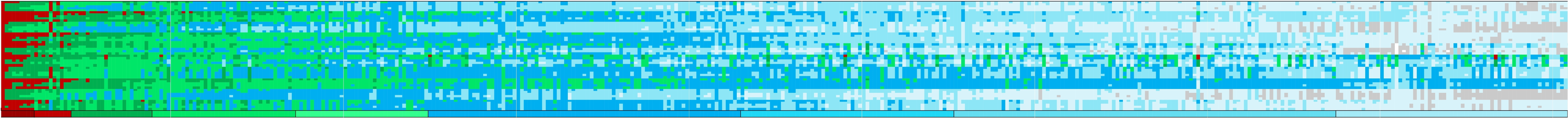}
\caption{\vvg{Color stratification of group membership ordered by the \textit{peer rank}  for $\mathcal{R}_{p} \leqslant 426$. Each color within the mosaic follows the thresholds of \jfm{Table} \ref{tab:groupiness}. Color bands with bold boundary lines at the bottom depict the groups in \jfm{Table} \ref{tab:group} ranging from $A$++ to $B-$}.}
\label{fig:OGamma}
\end{figure}

\vvg{
\textbullet \, The \textit{peer similarity} creates a league table with stratification of performance among different groups, but within a group all members are \textit{peers} with little deviation in performance among themselves. For the sake of comparison with established league tables, I will differentiate between \textit{peers} by the mean groupiness $\langle G \rangle$, leading to a continuous \textit{peer ranking} $\mathcal{R}_{p}$ (see \jfm{Table} \ref{tab:group}). 
}

\end{displayquote}

\subsection{Results}\label{sec:results}

I \vvg{have thus far} assign\vvg{ed} a methodology \vvg{that is solely focused on} subject rankings \vvg{through the variables $(\Lambda_{i},\mathcal{N}_{j})$ and the faculty density correction $\varphi$. As a result, the overall evaluation of all institutions} is simply \vvg{composed by} the extensive analysis of all subjects. \vvg{More specifically}, intensive \vvg{(the \textit{Leiden} parameter $\mathcal{L}$)} and extensive \vvg{(the output $\mathcal{O}$)} parameters were assigned to every subject for all institutions and make an extensive analysis of all subjects, each with the same \vvg{maximum score to remove bibliometric disparity among different subjects}. \vvg{Because the overall rank is obtained from the \textit{peer similarity} defined in eqs.~(\ref{eq:G1}-\ref{eq:G2}), the final \textit{peer ranking} $\mathcal{R}_{p}$ independs on how the intensive and extensive measures are computed. In fact, one could use the $h$-index as an alternative to both $(\mathcal{L},\mathcal{O})$ without \y{significant deviations in} the final result. However, \y{in addition of being time-dependent \citep{Mannella2013},} it is computationally burdensome to find the main campus contribution $\rho \in [0,1]$ with the h-index. The core criteria that delineates final groups of similarity and ultimately the \textit{peer ranking} is the stratification of classes defined in \jfm{Table} \ref{tab:groupiness}, whose stratification is clearly seen in \jfm{Figure} \ref{fig:OGamma}.}
\begin{table*}[b!]
  \centering
\begin{tabular}{lll|rrrrrrrrrrrrrrr}   
\toprule
\emph{\vvg{Class}} & \emph{Grade} & \emph{\vvg{Group}} &
 \emph{$\langle X \rangle$} &  \emph{$\langle Y \rangle$} &  \emph{$\langle \overline{\Lambda_{X}} \rangle$}  &  \emph{$\langle \overline{\Lambda_{Y}} \rangle$} &  \emph{$\langle \mathcal{C} \rangle$}  &  \emph{$\langle \mathcal{N}_{200} \rangle$}  &   \emph{$\sum U $}  &  \emph{$\textbf{ Countries}  $}  
\\
\midrule
\bf{Elite}       & \colorbox{Saulo1}{\, A++ \,} & 1, 1+ 1/2  & \vvg{44,300} & \vvg{47,900} & \vvg{614.4} & \vvg{622.5} & \vvg{653.8} & \vvg{72} & \textbf{9} & \textbf{3} \\
                 & \colorbox{Saulo11}{\, A+\,\,\,\,\,\,\,}  & 2          & \vvg{31,100} & \vvg{39,000} & \vvg{522.0} & \vvg{504.4} & \vvg{537.2} & \vvg{63} & \textbf{\vvg{10}}  & \textbf{\vvg{4}}   \\
\bf{World-Class} & \colorbox{Saulo2}{\, A \quad \,\,\,\,}   & 2 + 1/2  & \vvg{29,700} & \vvg{33,000} & \vvg{478.7}   &  \vvg{472.9} &  \vvg{500.5} & \vvg{62}  & \textbf{\vvg{22}}  & \textbf{\vvg{8}}   \\
                 & \colorbox{Saulo3}{\, A- \quad \,\,}  &  3 & \vvg{23,300} & \vvg{27,300} & \vvg{433.4} &  \vvg{423.4} &  \vvg{423.8} & \vvg{54}  & \bf{39}  & \textbf{\vvg{13}}   \\
                 & \colorbox{Saulo33}{\, A- -\,\,\,\,\,\,}&  3 + 1/2 & \vvg{17,800} & \vvg{24,000} & \vvg{391.3}  &  \vvg{388.4}& \vvg{348.1}& \vvg{45}  & \textbf{\vvg{36}}  & \textbf{\vvg{14}}   \\
\bf{Continental} & \colorbox{Saulo4}{\, B++ \,} & 4  & \vvg{13,200}  & \vvg{19,800} & \vvg{377.7}  &  \vvg{356.6}&  \vvg{292.6} & \vvg{36}  & \textbf{\vvg{85}}  & \textbf{\vvg{26}}   \\
                 & \colorbox{Saulo44}{\, B+ \,\,\,\,\,\,}  & 4 + 1/2  & \vvg{8,700}  & \vvg{16,400} & \vvg{366.6}   &  \vvg{332.2} &  \vvg{234.6}  & \vvg{24}  & \textbf{\vvg{58}}  & \textbf{\vvg{30}}   \\
                 & \colorbox{Saulo5}{\, B \quad \,\,\,\,}    & 5  & \vvg{6,300}  & \vvg{14,900} & \vvg{352.3}   &   \vvg{317.1} &  \vvg{191.9} & \vvg{18}  & \textbf{\vvg{104}} & \textbf{\vvg{32}}   \\
\bf{National}    & \colorbox{Saulo55}{\, B- \quad \,\,}  & 5 + 1/2  & \vvg{4,200}  & \vvg{11,600} & \vvg{361.6}   &   \vvg{274.9}&   \vvg{146.7} &  \vvg{12}  & \textbf{\vvg{63}}  & \textbf{\vvg{37}}   \\
                 & \colorbox{Saulo6}{\, B- -\,\,\,\,\,\,}& 6  & \vvg{2,700}  & \vvg{11,300} & \vvg{328.8}   &   \vvg{267.4}&   \vvg{111.0} &  \vvg{8}  & \textbf{\vvg{103}} & \textbf{\vvg{42}}   \\
\bf{Regional}    & \colorbox{Saulo7}{\, C+ \,\,\,\,\,}  & 6 + 1/2  &   \vvg{1,400}  & \vvg{9,400} & \vvg{326.4}  &   \vvg{233.5}&   \vvg{70.6} &  \vvg{4}  & \textbf{\vvg{160}} & \textbf{\vvg{47}}   \\
                 & \colorbox{Saulo77}{\, C \quad \,\,\,\,}   & 7  &   \vvg{700}  & \vvg{7,400} & \vvg{302.2}   &   \vvg{187.8} &   \vvg{42.4} &  2  & \textbf{\vvg{294}} & \textbf{\vvg{55}}   \\
\bf{Local}       & \colorbox{Saulo777}{\, D \quad \,\,\,\,}   & 7 + 1/2, 8  & \vvg{300} & \vvg{5,500} & \vvg{261.0}   & \vvg{155.5}  & \vvg{24.3} &  \vvg{1}   & \textbf{\vvg{270}} &  \textbf{\vvg{63}}   \\
\midrule                  
\bf{All}         &   &  & \vvg{4,900}  & \vvg{11,900}  & \vvg{304.9} & \vvg{250.5}   &   \vvg{127.6}&   \vvg{13} & \textbf{\vvg{1,253}}  & \textbf{\vvg{63}}  \\
\bottomrule
\end{tabular}
\caption{Group averages \vvg{$\langle \cdot \rangle $} among different criteria, the count of universities within each group and cumulative distribution of affiliated countries. Geographical distribution of group members allows the estimate of their influence (class).}
\label{tab:group}
\end{table*}
\begin{table*}
  \centering
\begin{tabular}{r|rrrrrrrrrrrrrrr}   
\toprule
\emph{Type} & 
 \emph{$Top1$} &   \emph{$Top10$} &  \emph{$Top30$} &  \emph{$Top100$}    &  \emph{$Top200$}  
\\
\midrule
\bf{Specialized}     &     \bf{0/80}   &  \bf{21/800} (2.6 \%)  &  \bf{56/2400} (2.3 \%) &  \bf{213/8000} (2.7 \%)  &  \bf{477/16000} (3.0 \%)  \\
\bottomrule
\end{tabular}
\caption{Total number of specialized institutions appearing in each \y{partition} of the subject rankings. \vvg{Nearly 180 of the total 1253 ranked institutions are specialized, representing 14\% of all evaluated universities.}}
\label{tab:group2}
\end{table*}
\begin{figure}
\begin{subfigure}[t]{0.44\textwidth}
    \includegraphics[scale=0.55]{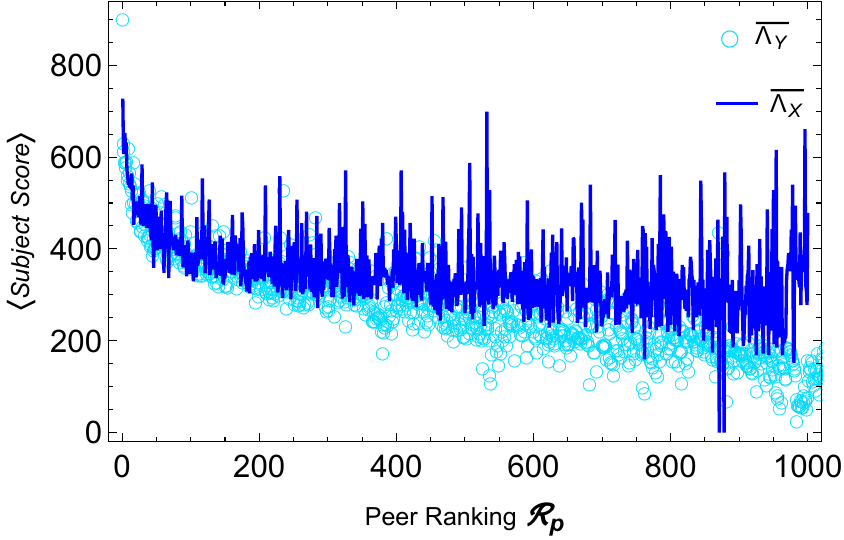}
\end{subfigure}
\hfill
\begin{subfigure}[t]{0.5\textwidth}
    \includegraphics[scale=0.57]{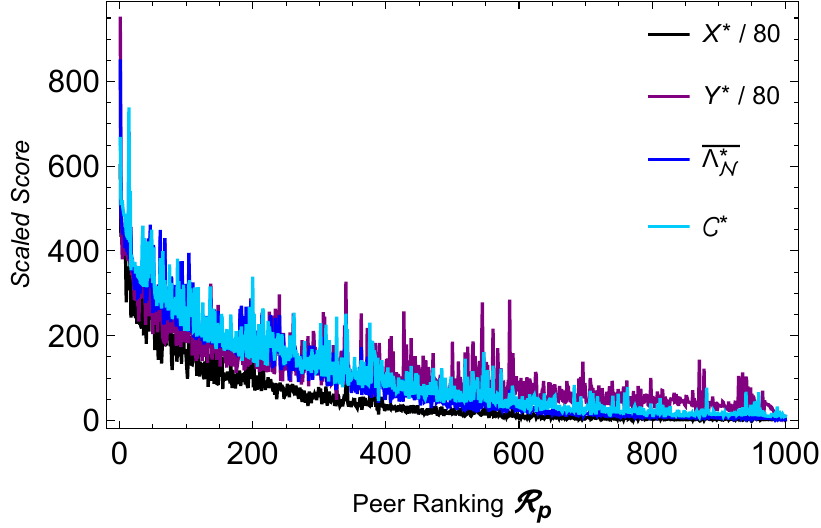}
\end{subfigure}
\caption{\vvg{Characteristic decay of the variables of the \textit{peer similarity} analysis: (a) mean subject score\y{s} and (b) absolute subject score\y{s} adjusted to faculty size.}}
\label{fig:SubjPeerRanking}
\end{figure}
\begin{figure}
\centering
    \includegraphics[scale=0.55]{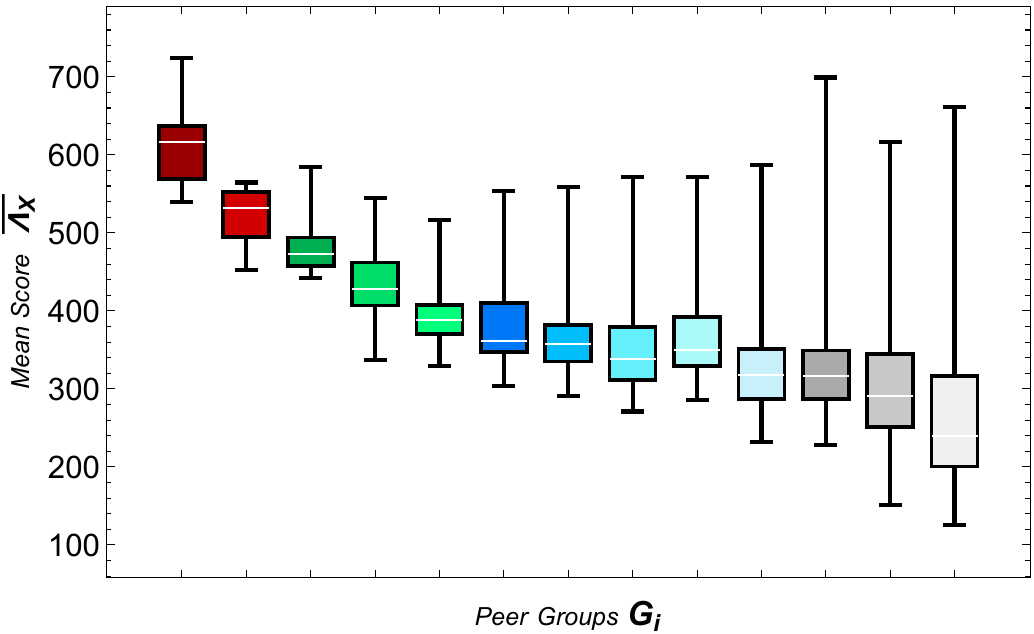}
\caption{\vvg{Box-whiskers diagram of the characteristic exponential decay of the mean subject score. This fast decay halts and reaches a region of stability at $200 \lesssim \mathcal{R}_{p} \lesssim 1250$ for the remaining groups. Color stratification is the same as in \jfm{Table} \ref{tab:group}.}}
\label{fig:Boxplot}
\end{figure}

\vvg{The main result is observed in \jfm{Figures} \ref{fig:SubjPeerRanking}\jfm{a} and \ref{fig:Boxplot}: the mean subject score $\overline{\Lambda_{X}}$ curve shows a clear stabilization in the tail ($\mathcal{R}_{p} \gtrsim 200$), establishing that almost all research-intensive universities attain the same \textit{quality}, i.e. the same average research impact in a number of academic subjects. The exception to such region of stability seems to be restricted to the first 1\y{0}0 ranked institutions. Conversely, \jfm{Figure} \ref{fig:SubjPeerRanking}\jfm{b} demonstrates \y{that} the remaining variables \y{are} strongly correlated with the \textit{peer ranking}, without \y{considerable} deviations among themselves. \y{Interestingly,} only a dozen countries appear in the groups of highest impact, representing 40\% of all subject rank slots (see \jfm{Table} \ref{tab:group}).}
\begin{figure}
\begin{subfigure}{0.33\textwidth}
    \includegraphics[scale=0.46]{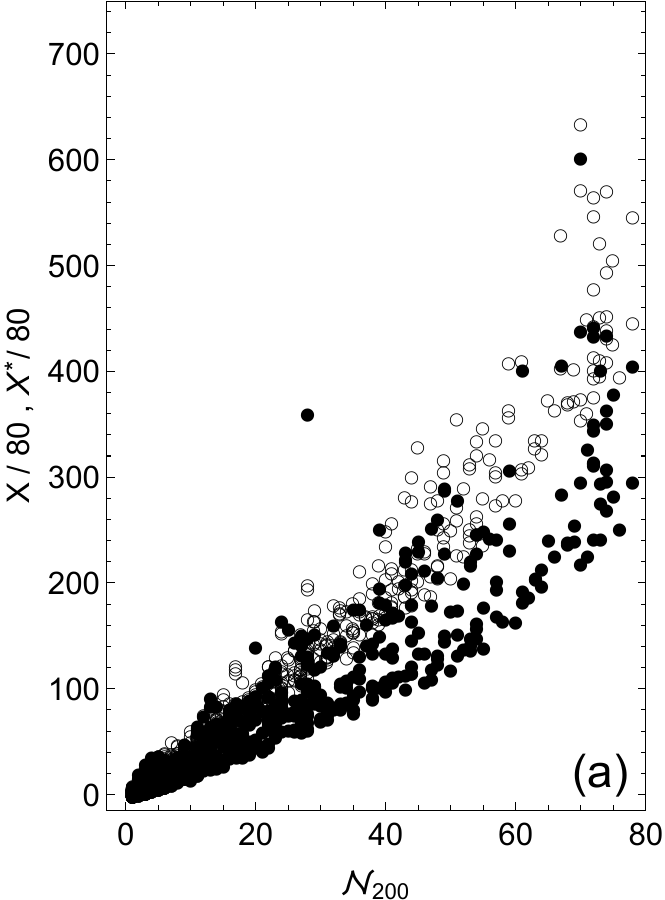}
\end{subfigure}
\begin{subfigure}{0.33\textwidth}
    \includegraphics[scale=0.46]{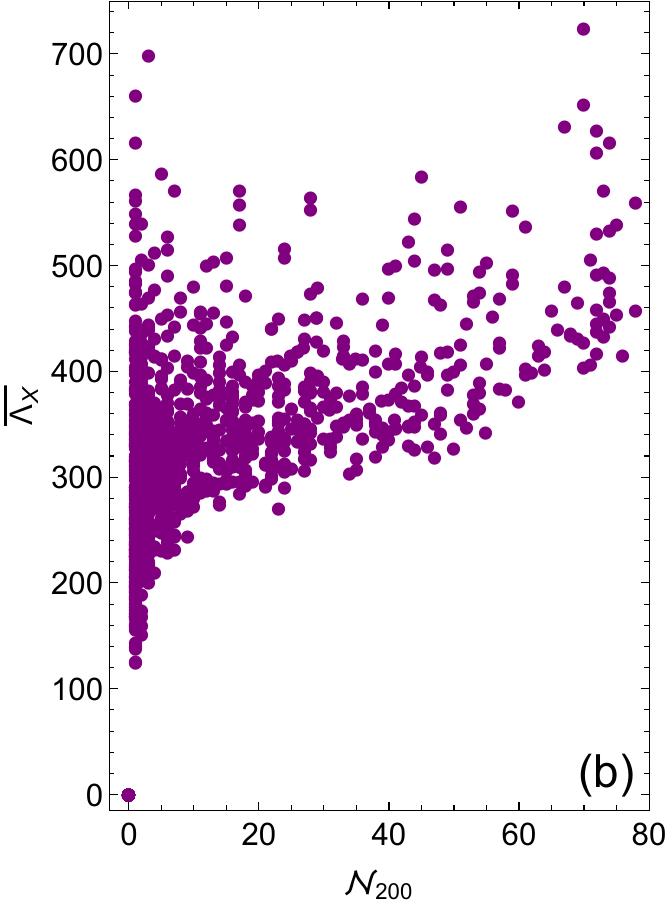}
\end{subfigure}
\begin{subfigure}{0.33\textwidth}
    \includegraphics[scale=0.46]{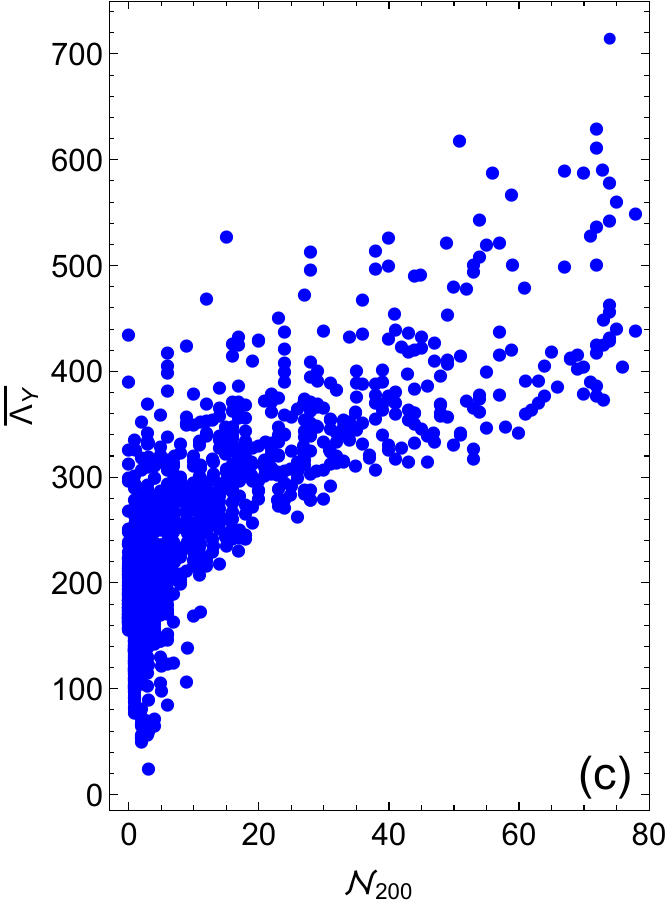}
\end{subfigure}
\caption{\vvg{Dependence on the total number of subjects among the top 200 ($\mathcal{N}_{200}$) for some key variables: (a) Total normalized subject score (circles) and the faculty size-adjusted equivalent (dots), (b) the \y{size-independent} mean subject score and (c) the \y{size-dependent} mean subject score.}}
\label{fig:SubjectN}
\end{figure}
\vvg{\y{Furthermore, w}e note in \jfm{Figure} \ref{fig:SubjectN}\jfm{a} that there is a strong correlation between the number of subjects exceled at and the normalized total subject score. This correlation is weakened when considered the \y{size-dependent} mean score per subject in \jfm{Figure} \ref{fig:SubjectN}\jfm{c}, while \jfm{Figure} \ref{fig:SubjectN}\jfm{b} demonstrates that no correlation exists \y{with} the \y{size-independent} mean score. \jfm{Figure} \ref{fig:SubjectN}\jfm{b} also reveals that excluding the institutions \y{with} $\mathcal{N}_{200}\geqslant 40$ the scatter plot of the mean score is identical to a normal distribution centered at $\overline{\Lambda_{X}} \approx 350$. This suggests that below the threshold  $\mathcal{N}_{200} = 40$ the mean subject score is simply randomic. Therefore, research-intensive institutions \y{seem to} mantain the same \textit{mean quality} (the reader should understand it as the mean research impact per field, that is to say, the degree of leadership at a given subject), except above a threshold which is the only feasible definition of a \textit{World-Class} institution.}

\vvg{Contrary to common belief, \jfm{Table} \ref{tab:group2} shows that specialized institutions do not tend to lead in their respective subjects, as they are underrepresented in the group of leading institutions in their areas of expertise. These institutions are considered specialized because they \y{do not} cover more than 20 subjects in terms of departamental structure. \y{In spite of their original names}, most technical universities and institutes do not limit their scope to technology and natural sciences, and thus can not be considered specialized institutions. For a comparison, major nominal technical institutions include \textit{MIT} with $\mathcal{N}_{200}=67$, \textit{ETH Zurich} \y{with} $\mathcal{N}_{200}=61$, and \textit{Technical Univ. Munich} \y{with} $\mathcal{N}_{200}=49$ whereas truly specialized institutions are represented by \textit{UC San Francisco} and the \textit{London School of Economics (LSE)} both with $\mathcal{N}_{200}=17$}.

\vvg{\y{Despite} the \textit{peer ranking} \y{being} a measure of homogeneity among institutions of each group \y{and} the thresholds not being dependent on comprehensiveness (see \jfm{section} \ref{sec:methods}), \jfm{Figure} \ref{fig:NGRp}\jfm{a} conveys a clear mathematical relationship \y{between the \textit{peer ranking} and comprehensiveness of the kind}:}
\begin{equation}
\vvg{    
\mathcal{N}_{200} \approx 80\,  \textrm{exp} \,  \left[ \frac{1-\mathcal{R}_{p}}{200} \right]  \quad \therefore \quad \mathcal{R}_{p} \approx 1 + 200 \ln{ \left(  \frac{80}{ \mathcal{N}_{200}}  \right)} \quad .
}
\label{eq:NRp}
\end{equation}
\vvg{According to eq.~(\ref{eq:NRp}), it can be stated with a high degree of confidence that only one strong predictor for the \textit{continuous} league table emerges, namely the institutional comprehensiveness. Moreover, \jfm{Figure} \ref{fig:NGRp}\jfm{b} reveals that the mean groupiness also displays an exponential relationship with the peer ranking. \y{Note that} the groupiness requires threshold definitions and robustness thereof (see \jfm{Table} \ref{tab:groupiness}), whereas the measure of comprehensiveness $\mathcal{N}_{200}$ is self-explanatory and independent of group thresholds. Nonetheless, the groupiness and comprehensiveness seem to be related almost exactly by a linear relationship, as shown in \jfm{Figure} \ref{fig:NGRp}\jfm{c}.}

\section{The \textit{World-Class} Debate}

\begin{figure}
\begin{subfigure}[t]{0.33\textwidth}
    \includegraphics[scale=0.42]{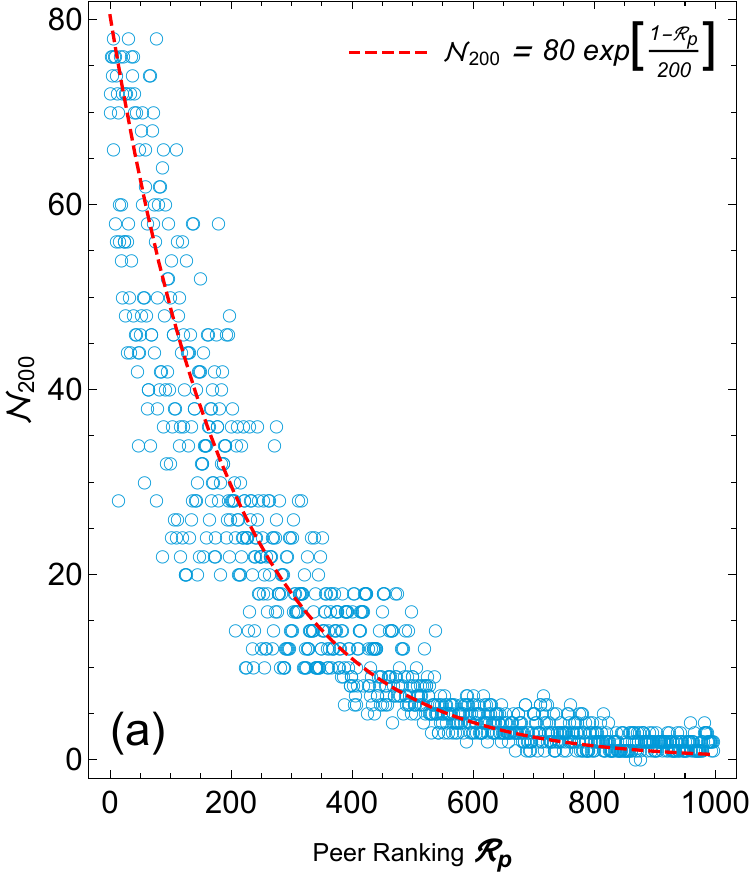}
\end{subfigure}
\begin{subfigure}[t]{0.33\textwidth}
    \includegraphics[scale=0.41]{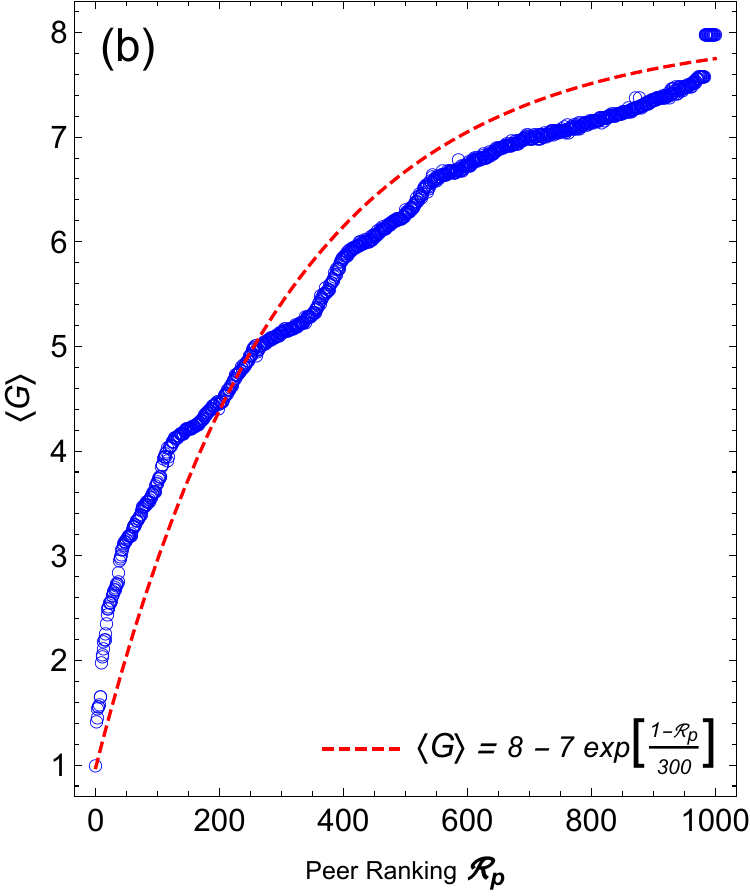}
\end{subfigure}
\begin{subfigure}[t]{0.33\textwidth}
    \includegraphics[scale=0.42]{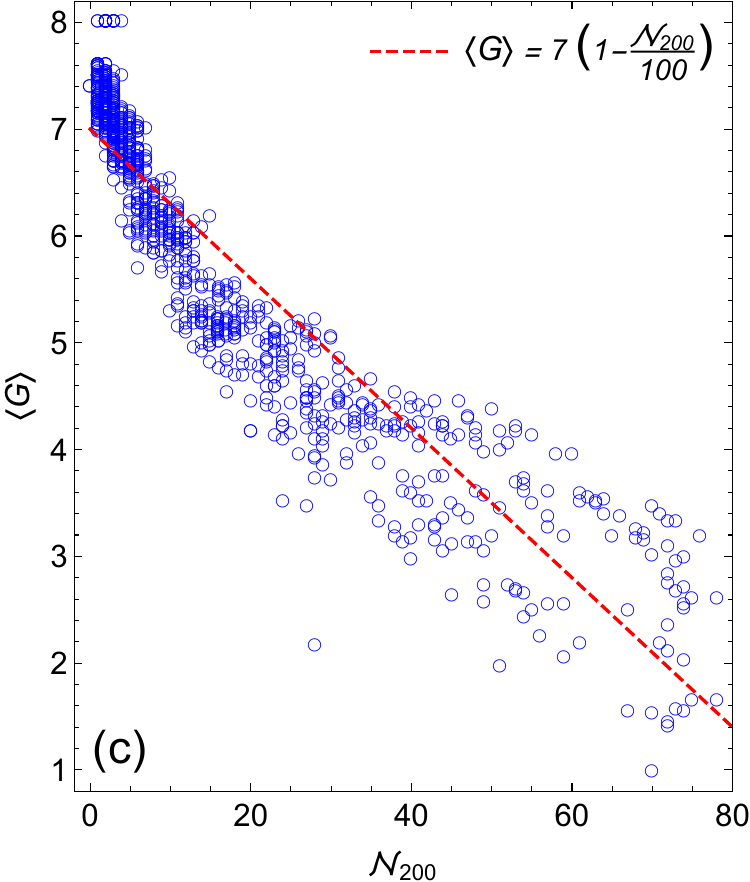}
\end{subfigure}
\caption{\vvg{Scatter plots between comprehensiveness $(\mathcal{N}_{200})$, \textit{peer ranking} $(\mathcal{R}_{p})$ and mean groupiness $(\langle G \rangle)$.}}
\label{fig:NGRp}
\end{figure}
\begin{figure*}[t]
\begin{subfigure}[t]{0.33\textwidth}
    \includegraphics[scale=0.46]{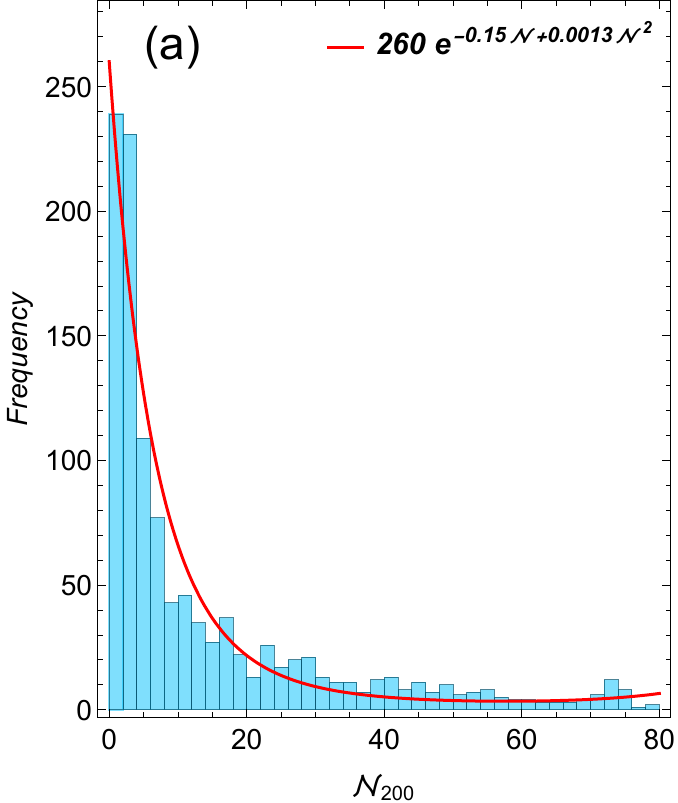}
\end{subfigure}
\begin{subfigure}[t]{0.33\textwidth}
    \includegraphics[scale=0.46]{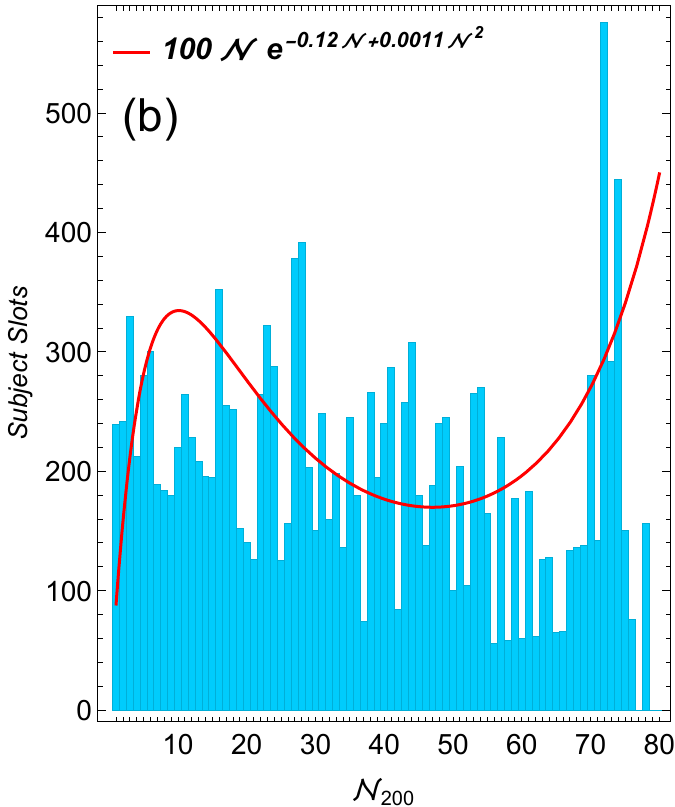}
\end{subfigure}
\begin{subfigure}[t]{0.33\textwidth}
    \includegraphics[scale=0.24]{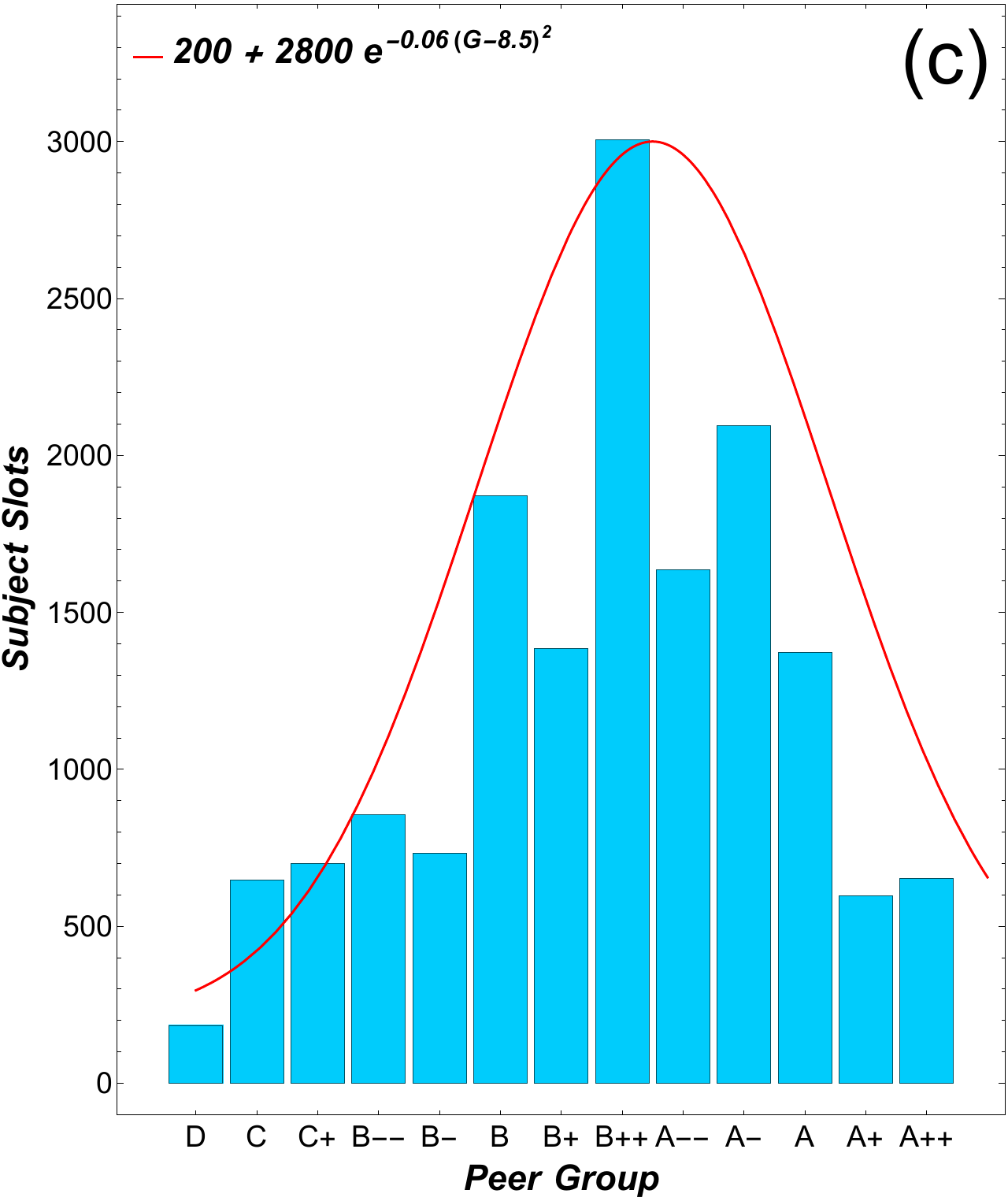}
\end{subfigure}
\caption{\vvg{Histograms and fitting curves (solid) for the (a) frequency of institutions with a given $\mathcal{N}_{200}$ (b) total number of slots in subject rankings distributed over $\mathcal{N}_{200}$ (c) and the mean groupiness $\langle G \rangle$.}}
\label{fig:histogram}
\end{figure*}

\y{Another important topic of debate regards whether is possible to reach a definition of world-class institutions} \citep{Hazelkorn2008}. \vvg{\y{C}ritics claim that strategic} national programs created to help their universities reach the \vvg{ill-defined} level of "world-class" institutions overlook th\y{e} major flaw of \textit{zero-sum game} \y{in} league tables, and it would be impossible to have all \y{or many of} these institutions defined as world-class \citep{Douglass2016}. \y{In this section I address this problem.} \vvg{First and foremost, it has been conclusively shown in \jfm{section} \ref{sec:results} that the significantly higher mean subject score is what distinguishes world-class institutions from the remaining ones. Secondly, I have also shown this group to be strongly correlated with comprehensiveness of research impact. Therefore, the present analysis has the capability of adding nuance to the debate: a world-class level can be achieved by institutions currently belonging to lower groups without affecting \y{its peers}, since the groupiness is strongly correlated with comprehensiveness (see \jfm{Figure} \ref{fig:NGRp}\jfm{c}). As a result, it is possible to enlarge the number of world-class institutions without affecting the degree of comprehensiveness \y{of the} remaining universities. This is possible because \y{of} the conservation principle:}
\begin{equation}
\vvg{
\int_{0}^{80} f_{s}(\mathcal{N}_{200}) \, d\mathcal{N}_{200} = 16,000 \quad ,
}
\end{equation}
\vvg{where $f_{s}(\mathcal{N}_{200})$ is the dimensional equivalent of a probability density counting the total number of subject ranking slots taken by all institutions with a given $\mathcal{N}_{200}$. As long as this integral property of the subject rankings holds, any redistribution of academic performance is possible. In fact, \jfm{Figure} \ref{fig:histogram}\jfm{a} shows that the vast majority of institutions belong to the lowest groups of \jfm{Table} \ref{tab:group}. However, in \jfm{Figure} \ref{fig:histogram}\jfm{b} I shift the focus to the histogram of subject ranking slots as a function of comprehensiveness, observing a median value of $\mathcal{N}_{200} \approx 37$  and thus the equal share of subject rank slots between the half bottom and top half groups. Nonetheless, the histogram is far from being uniformly distributed, displaying two maxima: first of the lowest groups ($\mathcal{N}_{200} \leqslant 30$) and the global maximum at the highest groups with $72 \leqslant \mathcal{N}_{200} \leqslant 74$. Similarly, \jfm{Figure} \ref{fig:histogram}\jfm{c} breaks down the same histogram into groupiness, revealing a normal distribution centered at the B++ group.} 

\vvg{\y{However,} the landscape of global research universities is changing quickly due to the rise of institutions from developing nations\y{, in particular from Asia} \y{\citep{veugelers2013world,veugelers2017challenge,index2021superpowered,conroy2022nature}}.} H\y{ence}, the current bibliometric trend will likely result in the convergence of the distribution of excellence in academic subjects to a uniform one in a few decades\y{,} decreasing the peak between groups B++ and A in \jfm{Figure} \ref{fig:histogram}\jfm{c} and reallocat\y{ing} them to neighboring lower groups. \vvg{Therefore, the \textit{peer similarity} approach removes the \textit{zero-sum game} issue:} although it is not possible for every developed/developing nation to claim 10 universities among the top 100, it is possible \vvg{\y{that the group of} world-class institutions \y{will grow}}. \vvg{Consequently, it is possible to} answer the questions \vvg{raised} by \citet{Altbach2011} and \citet{Oliver2013} on world-class institutions: Firstly, yes it is possible for most of the major 50 economies to have a few world-class universities. Secondly, the present statistical analysis extracted precise characteristics of this class.

The present methodology has shown that \vvg{continuously stratified} ranking tables are obsolete and should be replaced by similarity of university groups. Likewise, the Carnegie classification \citep{Carnegie1976,Kosar2018} puts together universities of different groups \vvg{of research intensity}, but the new model is more precise. \y{U}nlike \citet{Carnegie1976}, the model \y{is able to} differentiate between institutions like \textit{Caltech} and \textit{MIT}: while the former excels in 28 out of 80 subjects of \jfm{Table} \ref{tab:subjects} \vvg{and features a mean subject score of $\overline{\Lambda_{X}} \approx 560$}, the latter excels in 67 subjects \vvg{and reaches $\overline{\Lambda_{X}} \approx 640$}. Hence, the two universities can not be considered \vvg{peers and are not \y{in} the same tier}. \vvg{The \y{highest} Carnegie classification \y{(group), the} R1 institutions\y{, joins together} almost all \y{groups} of \jfm{Table} \ref{tab:group}.} The main disparity between the \citet{Carnegie1976} classification and \vvg{the present approach} is that \vvg{the former} merges all the research output of a university, \vvg{with} disproportionate allocation of research funds and efforts into a few subjects \vvg{of higher impact (see \jfm{Figure} \ref{fig:SubjectSeries})} compensat\vvg{ing} for the lack of it in the remaining subjects. Hence, in a comparison of universities, this methodology will not assess the uneven distribution of research output among subjects.

\section{Comparison with other league tables}\label{sec:projec}

\begin{figure}
\begin{subfigure}[t]{0.33\textwidth}
    \includegraphics[scale=0.42]{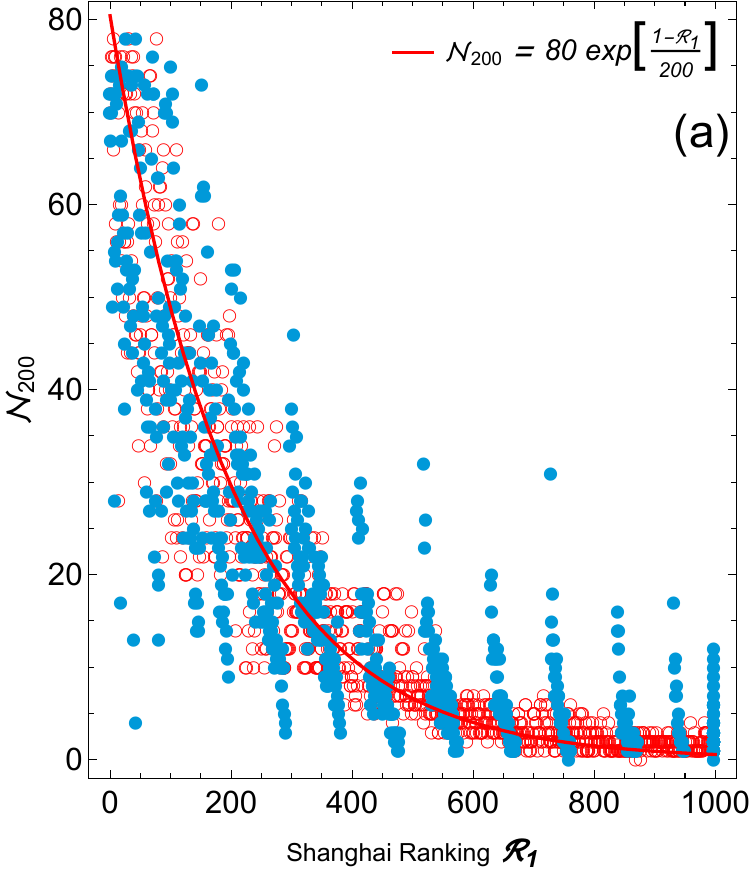}
\end{subfigure}
\begin{subfigure}[t]{0.33\textwidth}
    \includegraphics[scale=0.42]{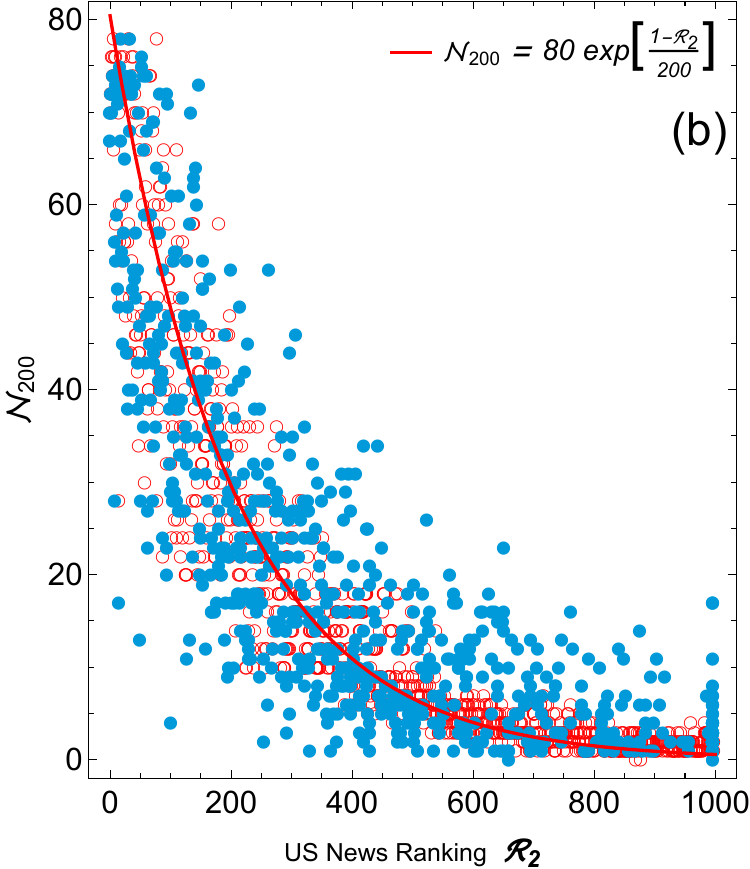}
\end{subfigure}
\begin{subfigure}[t]{0.33\textwidth}
    \includegraphics[scale=0.42]{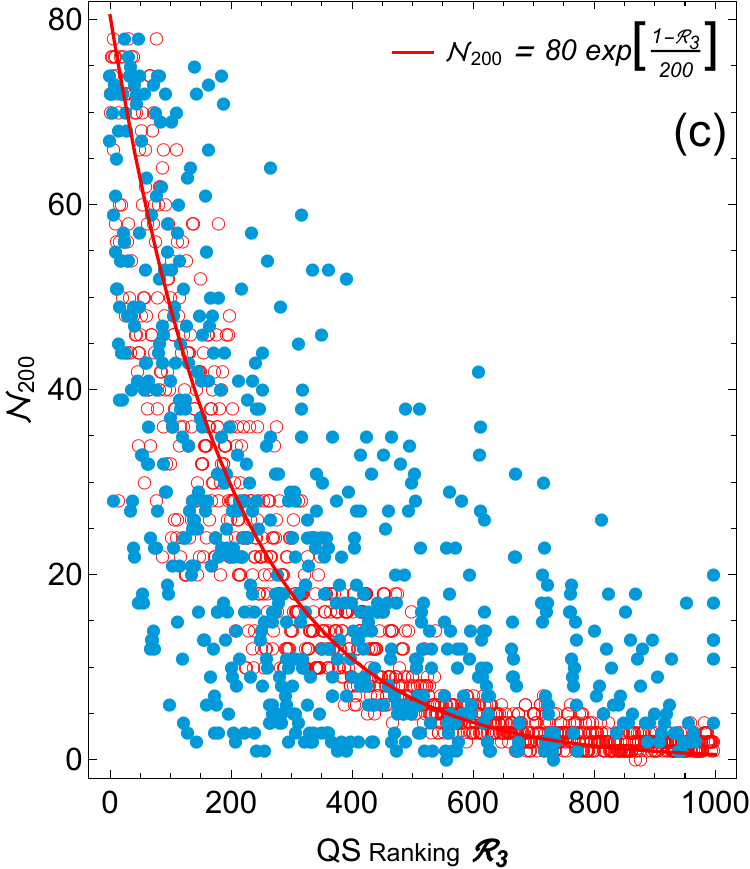}
\end{subfigure}

\begin{subfigure}[t]{0.33\textwidth}
    \includegraphics[scale=0.42]{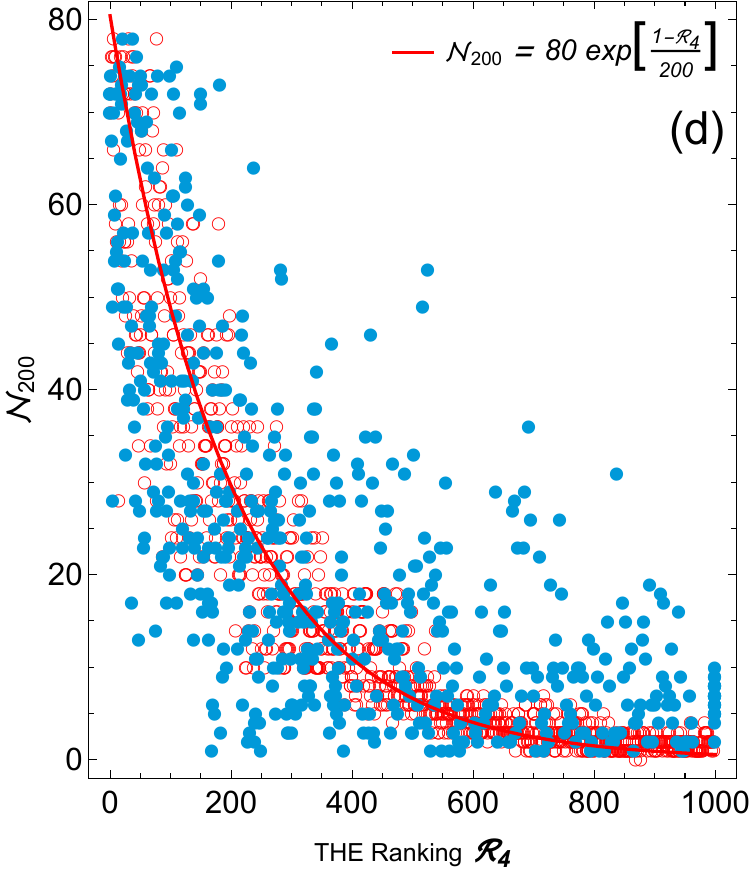}
\end{subfigure}
\begin{subfigure}[t]{0.33\textwidth}
    \includegraphics[scale=0.42]{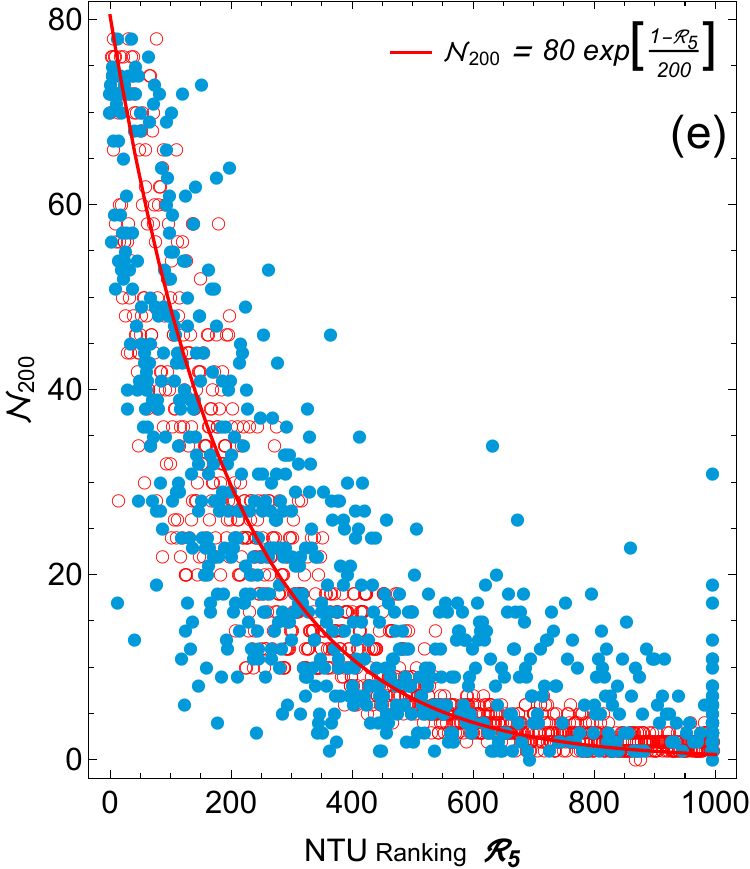}
\end{subfigure}
\begin{subfigure}[t]{0.33\textwidth}
    \includegraphics[scale=0.42]{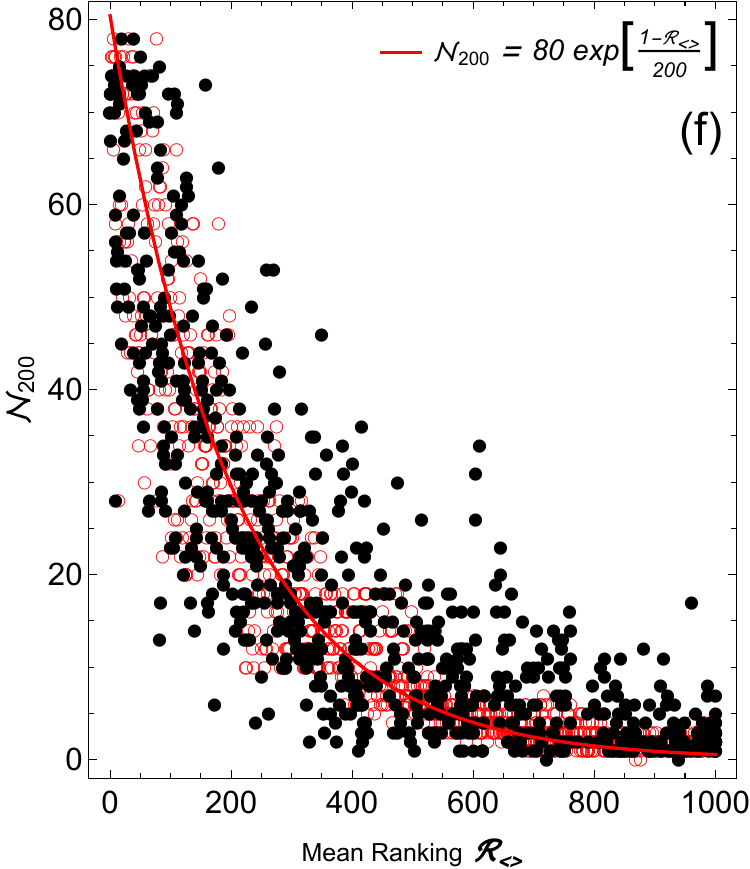}
\end{subfigure}
\caption{\vvg{Equivalent of \jfm{Figure} \ref{fig:NGRp}\jfm{a} for each institutional rank of the five most trusted ranking tables (dots) in their latest publications (2022) versus the underlying peer rank curve (red circles \y{and solid line}).}}
\label{fig:NGRp2}
\end{figure}
Even though the methodology of the five major university league tables are divergent, their core criteria are research impact. Hence, in this section I shall probe the main results of eq.~(\ref{eq:NRp}) as well as the mean subject score evolution in \jfm{Figure} \ref{fig:SubjPeerRanking}\jfm{a} against the five major university league tables in an attempt to reveal hidden features of their methodologies. First, I analyze how well the curve in \jfm{Figure} \ref{fig:NGRp}\jfm{a} is reproduced in established league tables. As can be observed in \jfm{Figure} \ref{fig:NGRp2}\jfm{a}, the \textit{Shanghai} ranking reproduces the same curve quite well. Note, however, that differences as compared to the peer ranking are visible: the Shanghai counterpart shows discrete stratification for $\mathcal{R}_{1} > 200$ (see \jfm{Table} \ref{tab:distRANK} for definitions),
which is due to their methodological discrete ranking above this threshold. Moreover, the \textit{Shanghai} ranking also features a few dozens outliers compared to the peer ranking. Likewise, the \textit{US News} league table follows the same curve of the peer rank, as shown in \jfm{Figure} \ref{fig:NGRp2}\jfm{b}. Although the correlation between both \textit{QS} and \textit{THE} rankings (see \jfm{Figures} \ref{fig:NGRp2}\jfm{c,d}) and the number of subjects has a much stronger scatter than the peer ranking, the trend of better ranked institutions to be more comprehensive is still observed. Finally, both curves for the \textit{NTU} and \textit{Mean} rankings (see \jfm{Figures} \ref{fig:NGRp2}\jfm{e,f}) are in good agreement with eq.~(\ref{eq:NRp}).

\begin{figure}
\centering
    \includegraphics[scale=0.5]{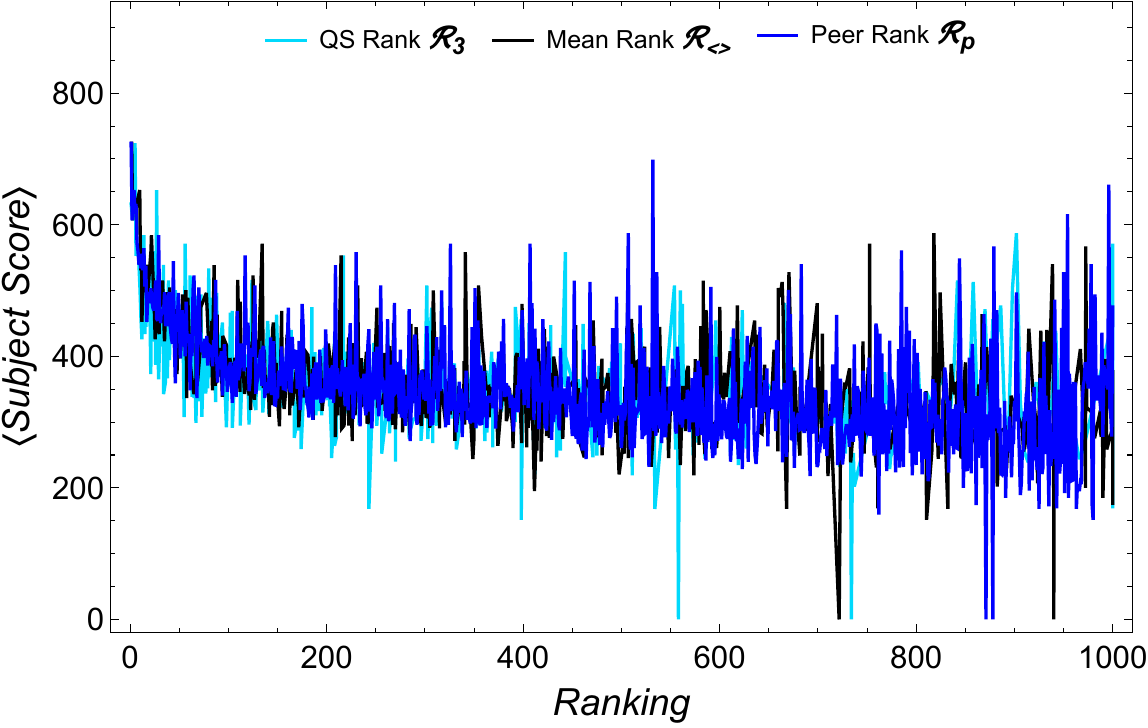}
\caption{\vvg{Equivalent (black curve) of \jfm{Figure} \ref{fig:SubjPeerRanking}\jfm{a} (blue curve) for the \textit{Mean} ranking among the five established league tables shown in \jfm{Figure} \ref{fig:NGRp2}}.}
\label{fig:qualitycomp}
\end{figure}

Furthermore, \jfm{Table} \ref{tab:distRANK} delineates by how much each institution rank differs between each league table and the peer rank. I found that on average, the vast majority of institutions lie in the $\pm 30\%$ confidence interval of the peer rank. This \y{interval supports} the view that institutions can not be 
continuously ranked, as the differences between peers \y{are} so small that their positions may change depending on each parameter weight \y{in} the overall computation. \y{Thus,} \jfm{Figure} \ref{fig:NGRp2} and \jfm{Table} \ref{tab:distRANK} confirm the validity of eq.~(\ref{eq:NRp}), and \y{allows us to} discern between the established league tables: \textit{Shanghai} and \textit{US News} are the most accurate while \textit{QS} and \textit{THE} are the least trustworthy.
\begin{table*}
  \centering
\begin{tabular}{r|rrrrrrrrrrrrrrr}   
\toprule
\emph{Range} & 
 \emph{$Shanghai (\mathcal{R}_{1})$} &   \emph{$USNews (\mathcal{R}_{2})$} &  \emph{$QS (\mathcal{R}_{3})$} &  \emph{$THE (\mathcal{R}_{4})$}    &  \emph{$NTU (\mathcal{R}_{5})$}  &  \emph{$Mean (\mathcal{R}_{<>})$}  
\\
\midrule
$0.90 \leqslant r \leqslant 1.10$ ; $\mathcal{R}_{p} \leqslant 1000$  & \bf{32.5\%} & \bf{26.6\%}  & \bf{16.7\%} & \bf{20.1\%} & \bf{25.9\%} & 28.0\% \\
$0.75 \leqslant r \leqslant 1.25$ ; $\mathcal{R}_{p} \leqslant 1000$  & \bf{64.7\%} & \bf{57.8\%}  & \bf{33.1\%} & \bf{43.1\%} & \bf{54.7\%} & 59.3\% \\
$0.50 \leqslant r \leqslant 1.50$ ; $\mathcal{R}_{p} \leqslant 1000$  & \bf{87.5\%} & \bf{84.7\%}  & \bf{59.9\%} & \bf{68.3\%} & \bf{78.8\%} & 84.5\% \\
\midrule
$0.90 \leqslant r \leqslant 1.10$ \,\, ;  $\mathcal{R}_{p} \leqslant 100$  & \bf{26.0\%} & \bf{30.0\%}  & \bf{13.0\%} & \bf{31.0\%} & \bf{29.0\%} & 25.0\% \\
$0.75 \leqslant r \leqslant 1.25$ \,\, ;  $\mathcal{R}_{p} \leqslant 100$  & \bf{61.0\%} & \bf{65.0\%}  & \bf{31.0\%} & \bf{53.0\%} & \bf{63.0\%} & 62.0\% \\
$0.50 \leqslant r \leqslant 1.50$ \,\, ;  $\mathcal{R}_{p} \leqslant 100$  & \bf{79.0\%} & \bf{88.0\%}  & \bf{56.0\%} & \bf{74.0\%} & \bf{79.0\%} & 86.0\% \\
\bottomrule
\end{tabular}
\caption{\vvg{University league table accuracy relative to the peer ranking $\mathcal{R}_{p}$ through their ratio $r$. The \textit{mean rank} measures the geometrical average among the five established league tables listed above.}}
\label{tab:distRANK}
\end{table*}
Additionally, I have also checked whether the distribution of mean score $\overline{\Lambda_{X}}$ as a function of the peer ranking also applies to other league tables. It could be that other league tables assign \y{so} different rank positions compared to $\mathcal{R}_{p}$ that the distribution of $\overline{\Lambda_{X}}$ would not saturate. As it turns out, \jfm{Figure} \ref{fig:qualitycomp} corroborates the main result that the mean score of universities across eighty academic subjects across all fields tend to saturate rather quickly outside the Elite and World-class groups (see \jfm{Table} \ref{tab:group}). Remarkably, even the curve for the \textit{QS} ranking shows no significant deviation from \jfm{Figure} \ref{fig:SubjPeerRanking}\jfm{a}, despite the former being the least accurate league table. 

\vvg{This assessment of the correlation between \y{the} \textit{peer rank} and these major league tables have have shown surprisingly good fidelity of \vvg{the measure of comprehensiveness. As such, it allows me to draw the conclusion that regardless of the exact criteria employed, the \y{continuous} league position of research-intensive institutions are strongly correlated with academic comprehensiveness. More precisely, the success of institutions in university league tables are strongly influenced by research comprehensiveness and secondarily by high density of score per faculty. 
The very few institutions, such as \textit{Caltech}, that are not fully comprehensive (but also not specialized) can appear among the best when their score density per faculty is very high. 
Conversely, a high score per faculty density can not uplift specialized institutions into groups of comprehensive ones, such as in the case of \textit{Scuola Normale Superiore di Pisa}.}}
\begin{figure}
\begin{subfigure}[t]{0.33\textwidth}
    \includegraphics[scale=0.3]{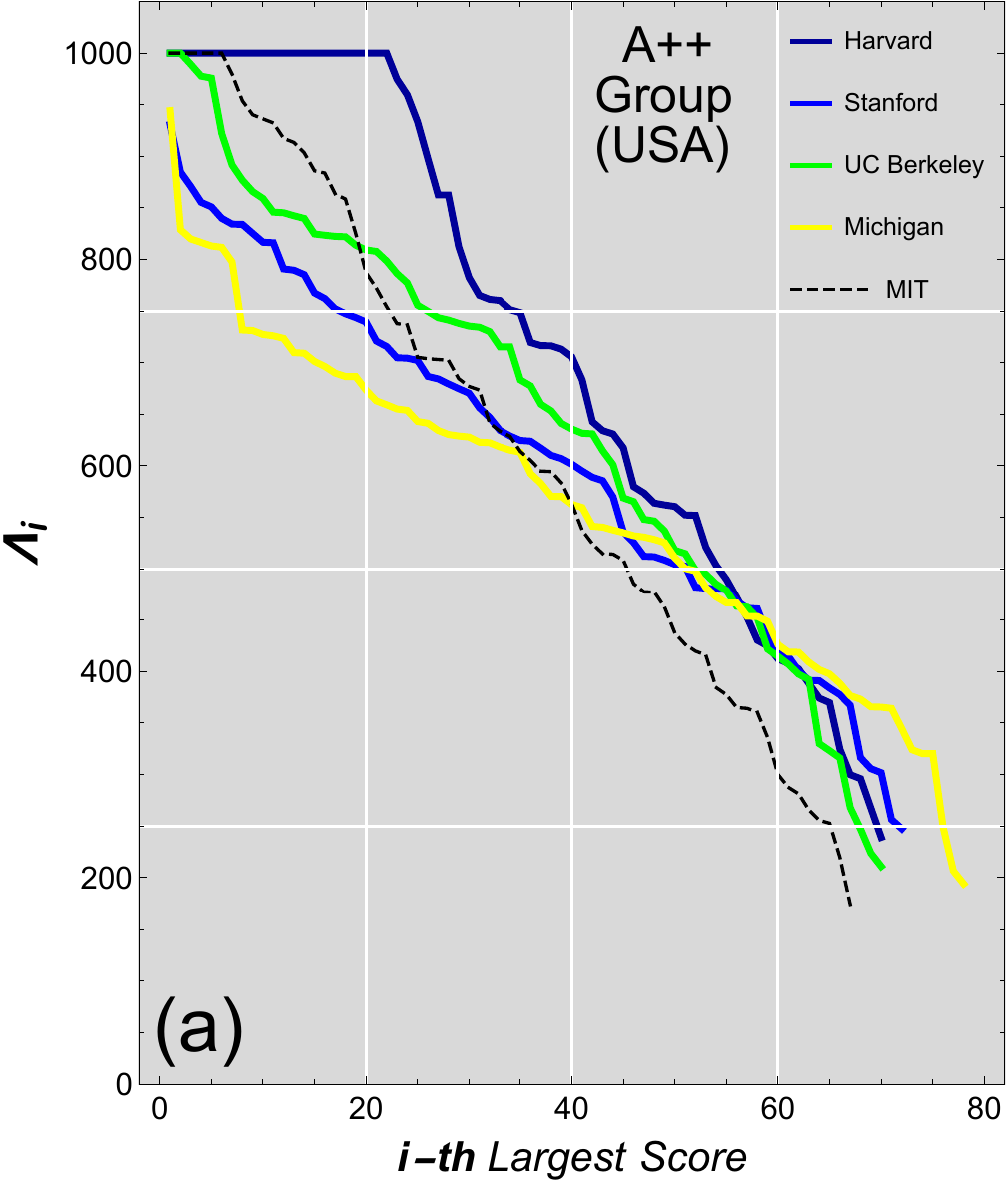}
\end{subfigure}
\begin{subfigure}[t]{0.33\textwidth}
    \includegraphics[scale=0.295]{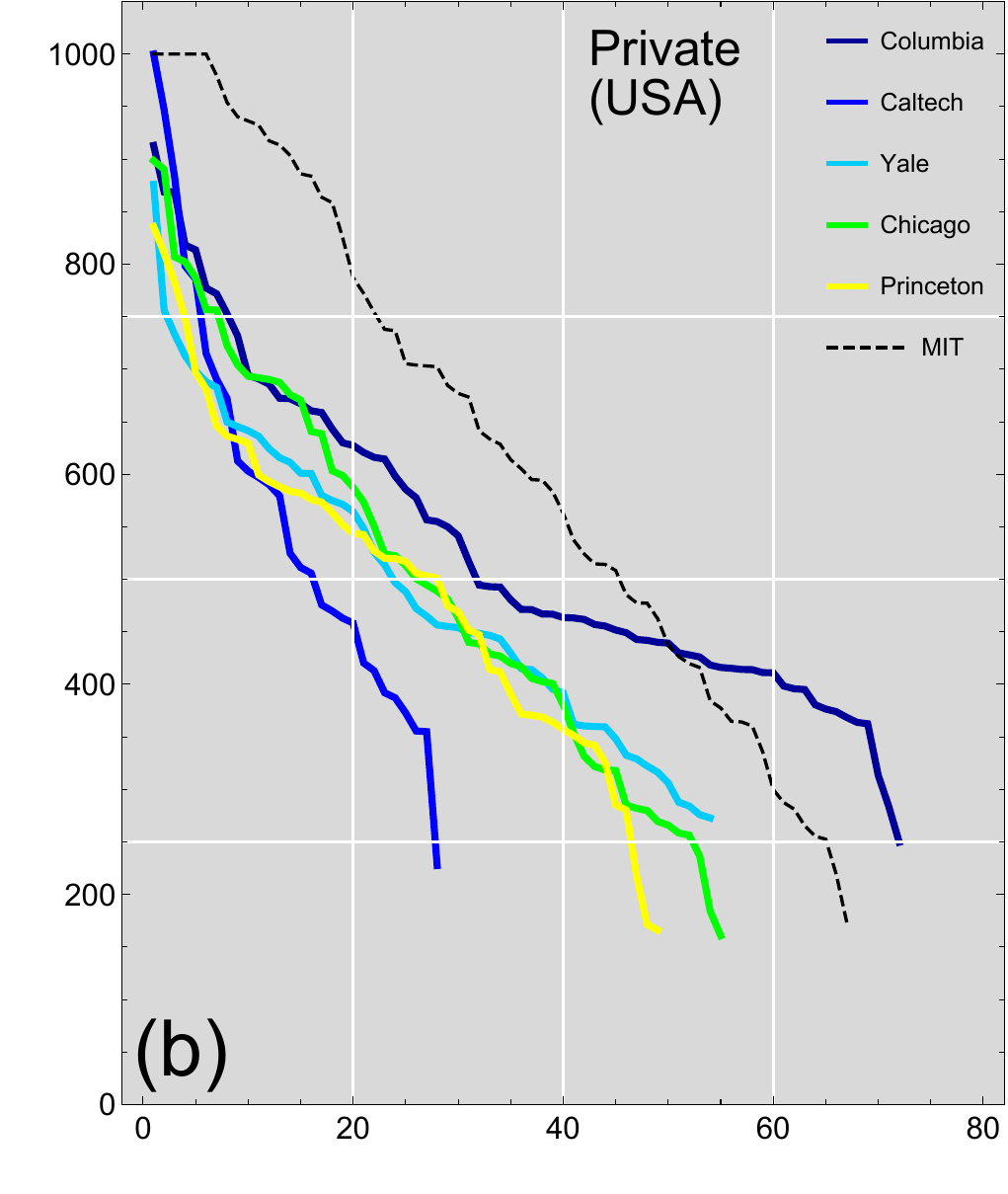}
\end{subfigure}
\begin{subfigure}[t]{0.33\textwidth}
    \includegraphics[scale=0.29]{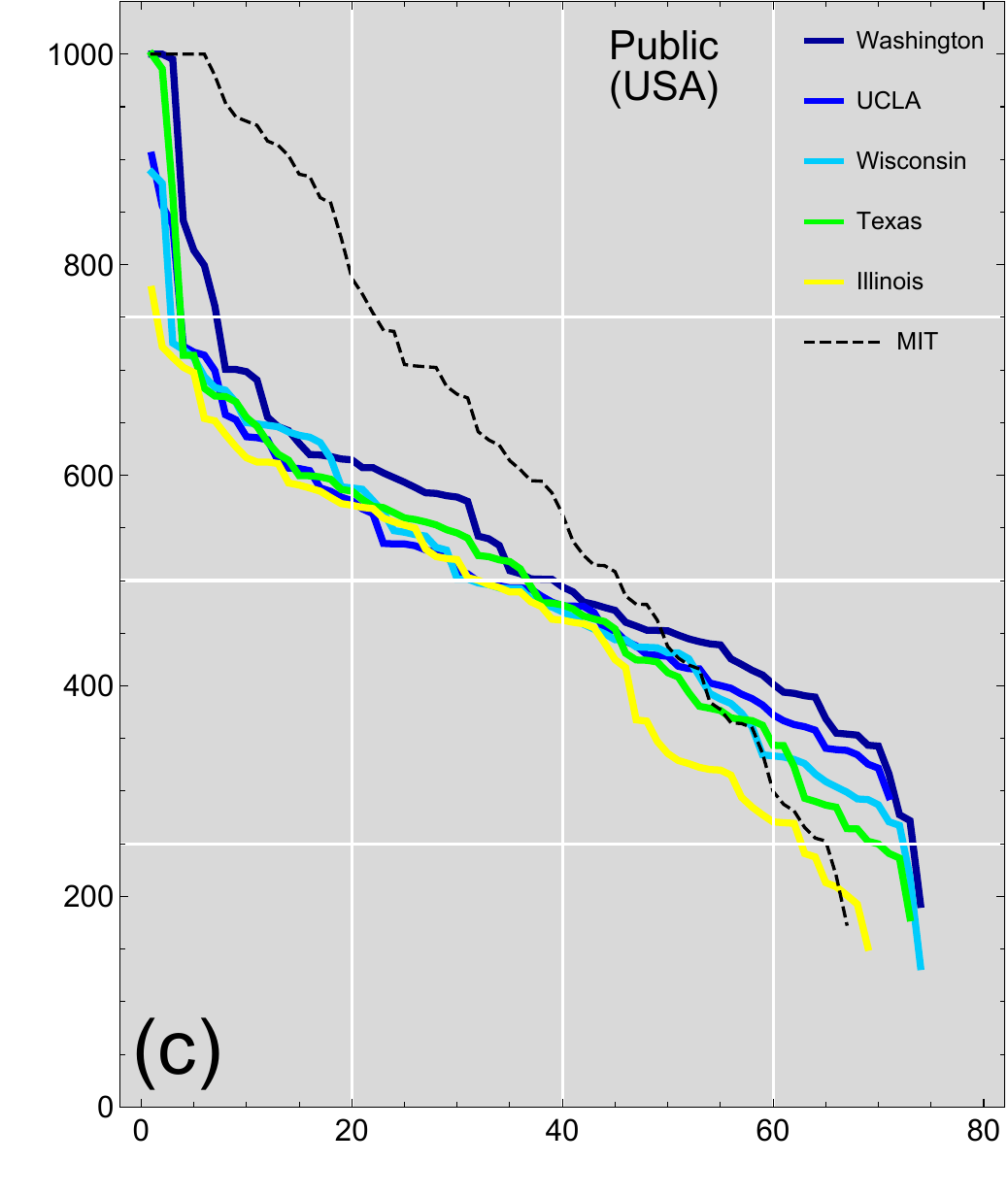}
\end{subfigure}

\begin{subfigure}[t]{0.33\textwidth}
    \includegraphics[scale=0.3]{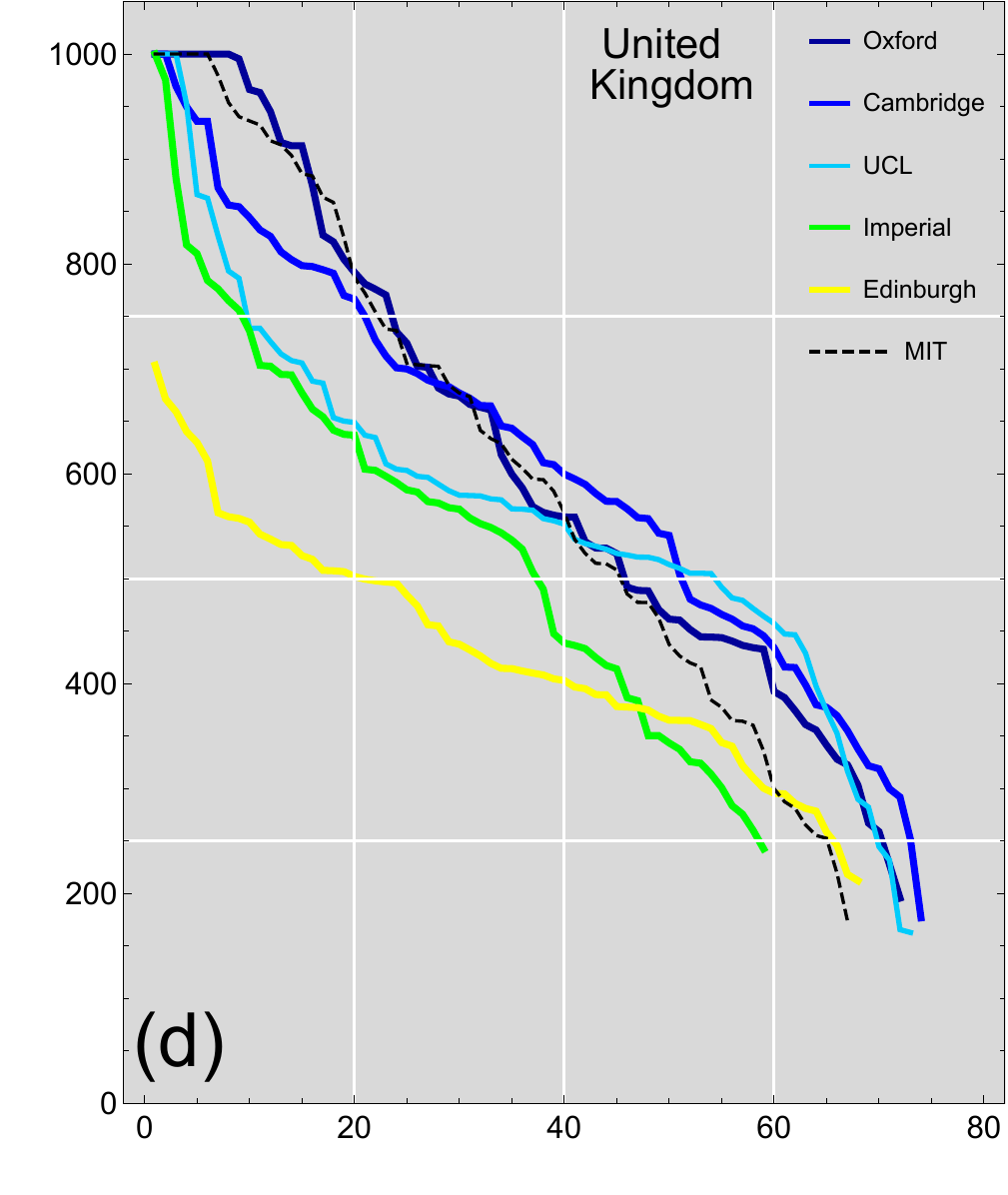}
\end{subfigure}
\begin{subfigure}[t]{0.33\textwidth}
    \includegraphics[scale=0.3]{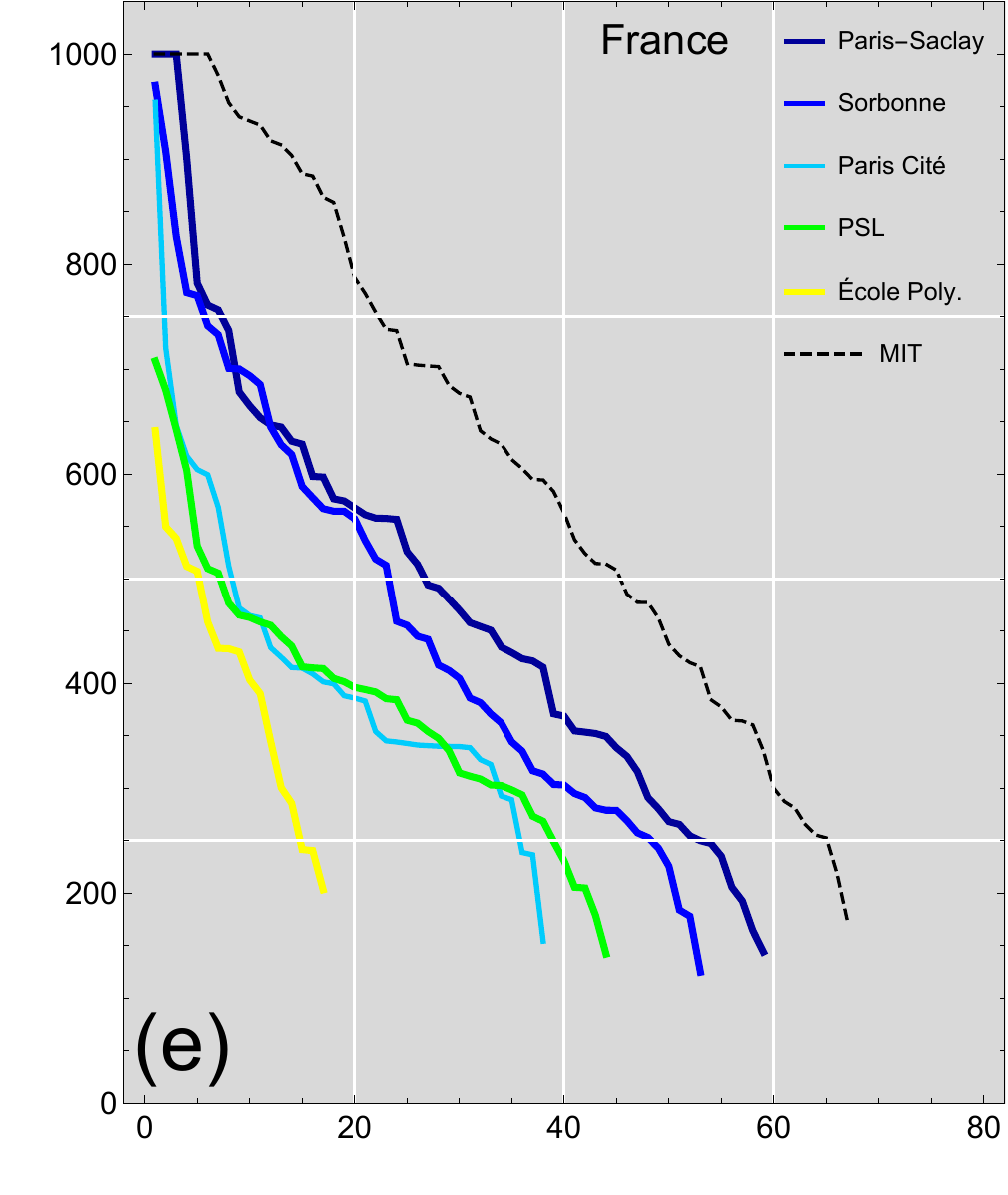}
\end{subfigure}
\begin{subfigure}[t]{0.33\textwidth}
    \includegraphics[scale=0.3]{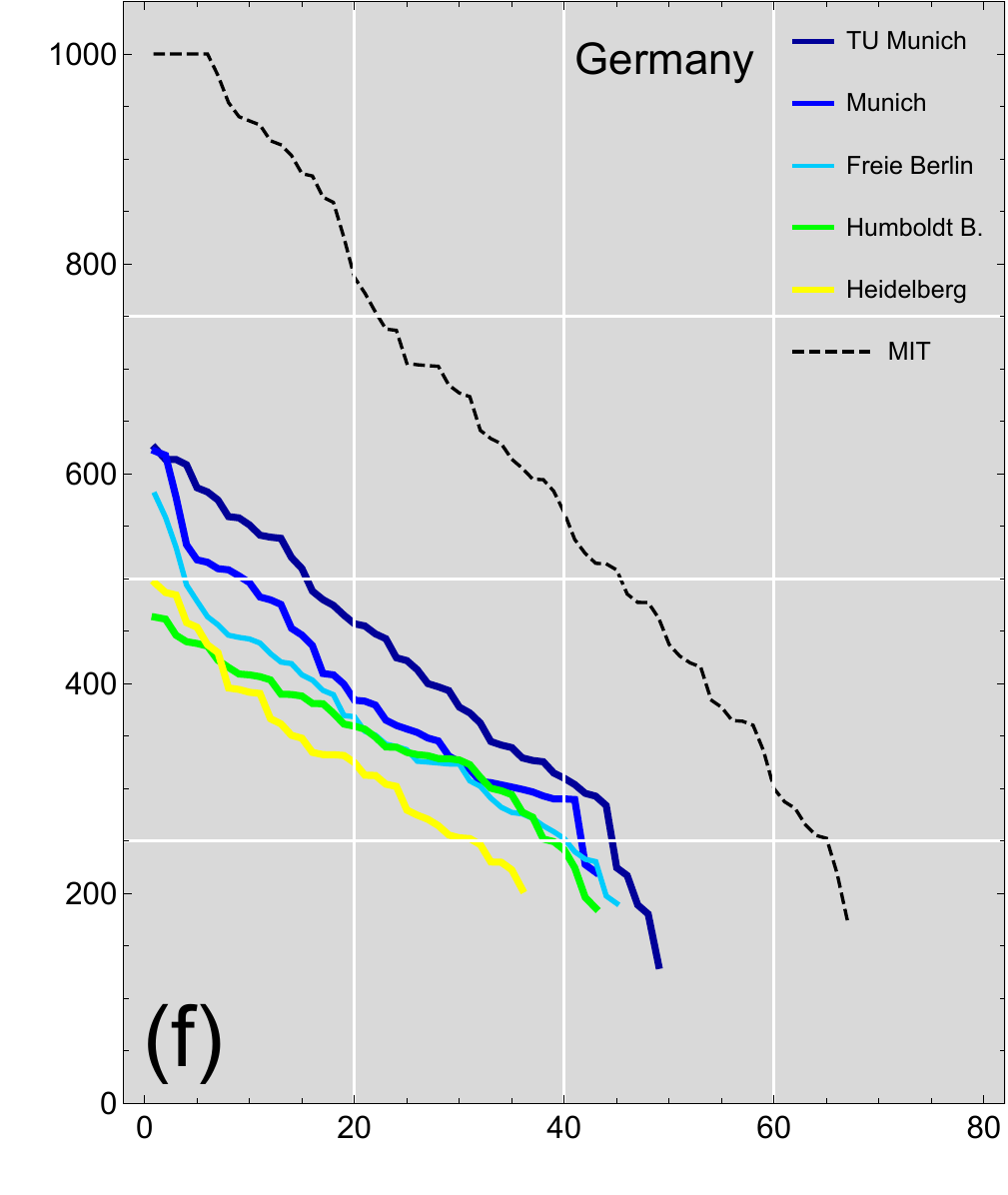}
\end{subfigure}

\begin{subfigure}[t]{0.33\textwidth}
    \includegraphics[scale=0.3]{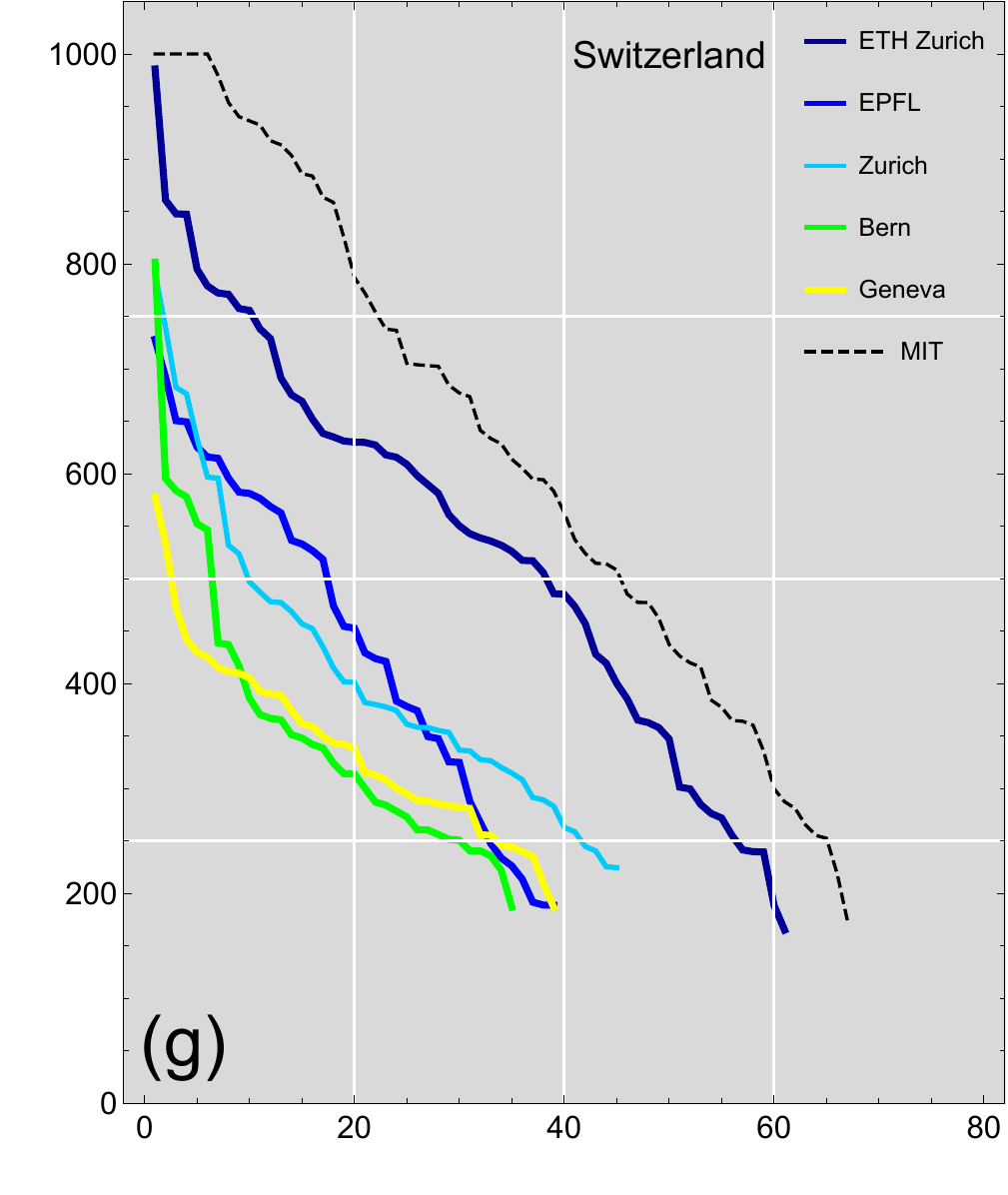}
\end{subfigure}
\begin{subfigure}[t]{0.33\textwidth}
    \includegraphics[scale=0.3]{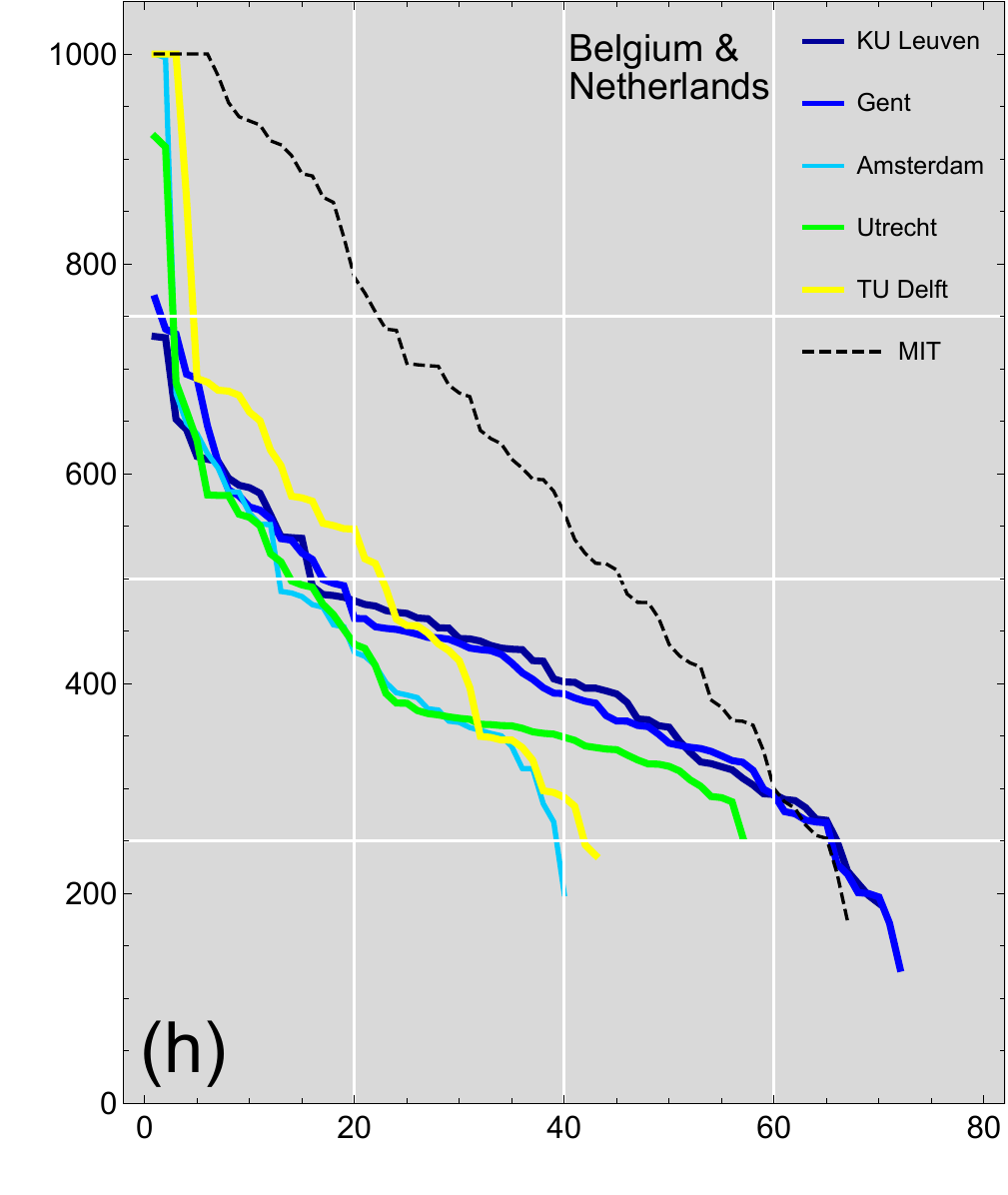}
\end{subfigure}
\begin{subfigure}[t]{0.33\textwidth}
    \includegraphics[scale=0.3]{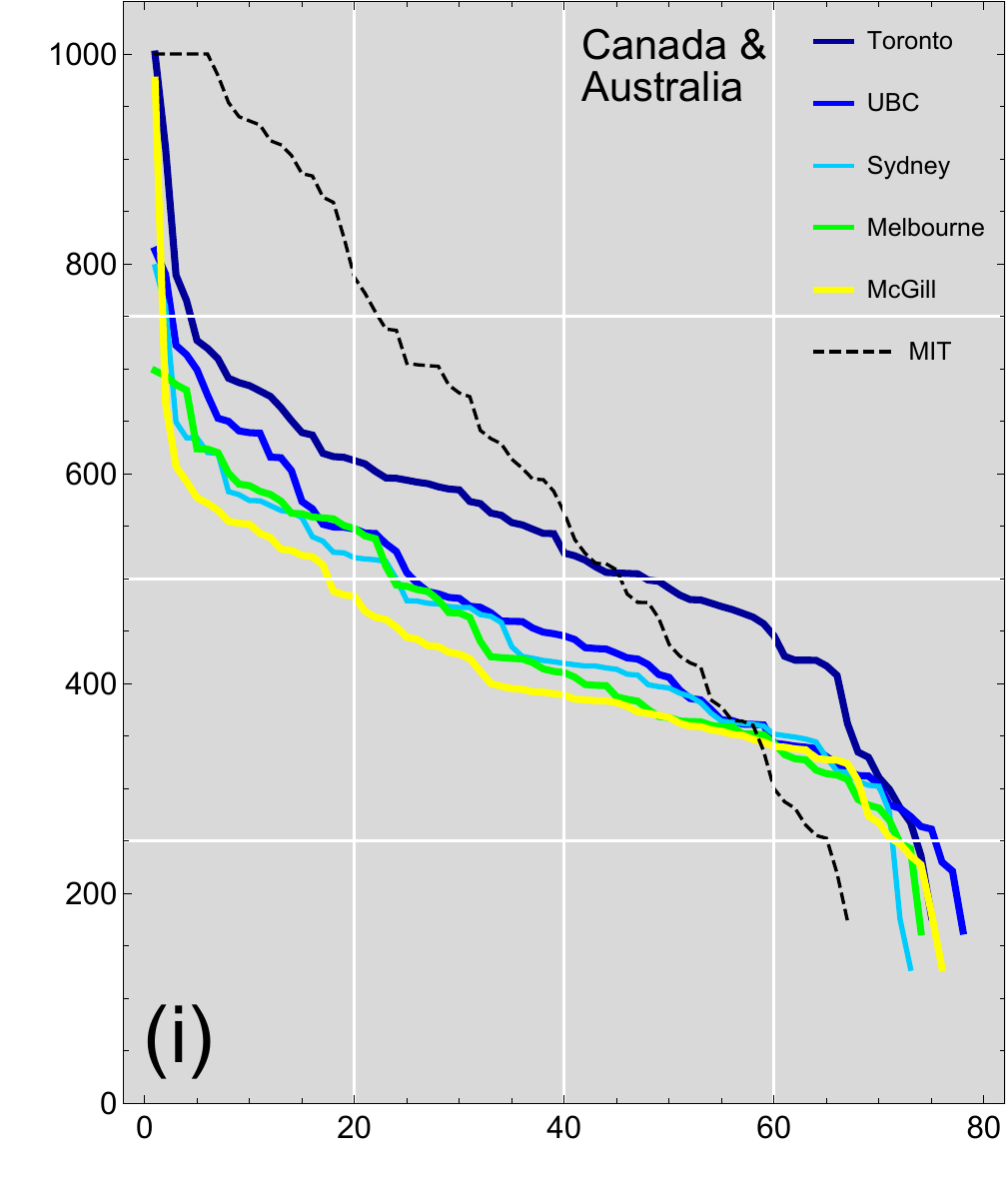}
\end{subfigure}
\caption{\vvg{Distribution of subject score order from highest to lowest of major institutions of leading countries in research output in the western hemisphere}.}
\label{fig:down1}
\end{figure}
\begin{figure}
\begin{subfigure}[t]{0.33\textwidth}
    \includegraphics[scale=0.3]{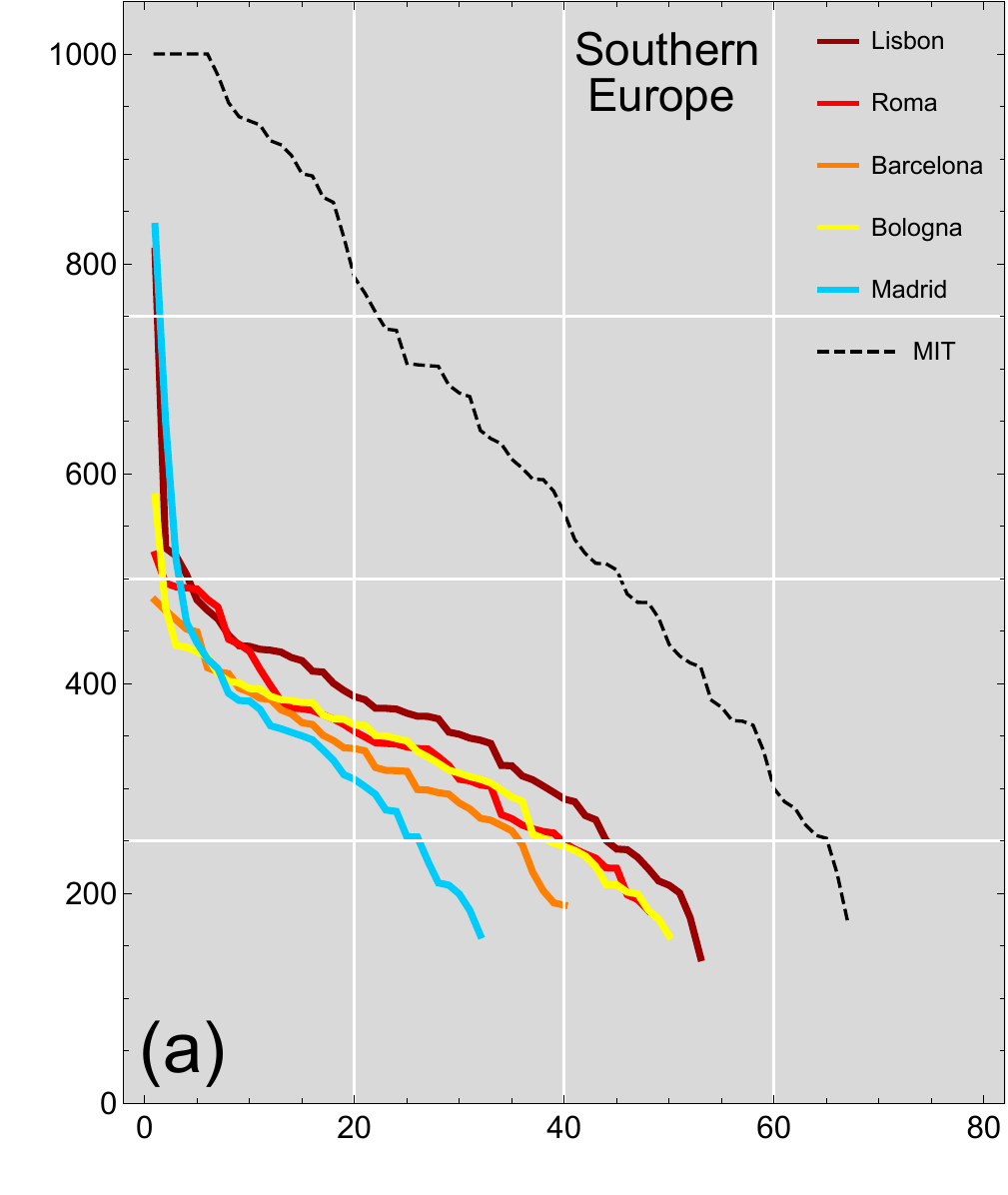}
\end{subfigure}
\begin{subfigure}[t]{0.33\textwidth}
    \includegraphics[scale=0.295]{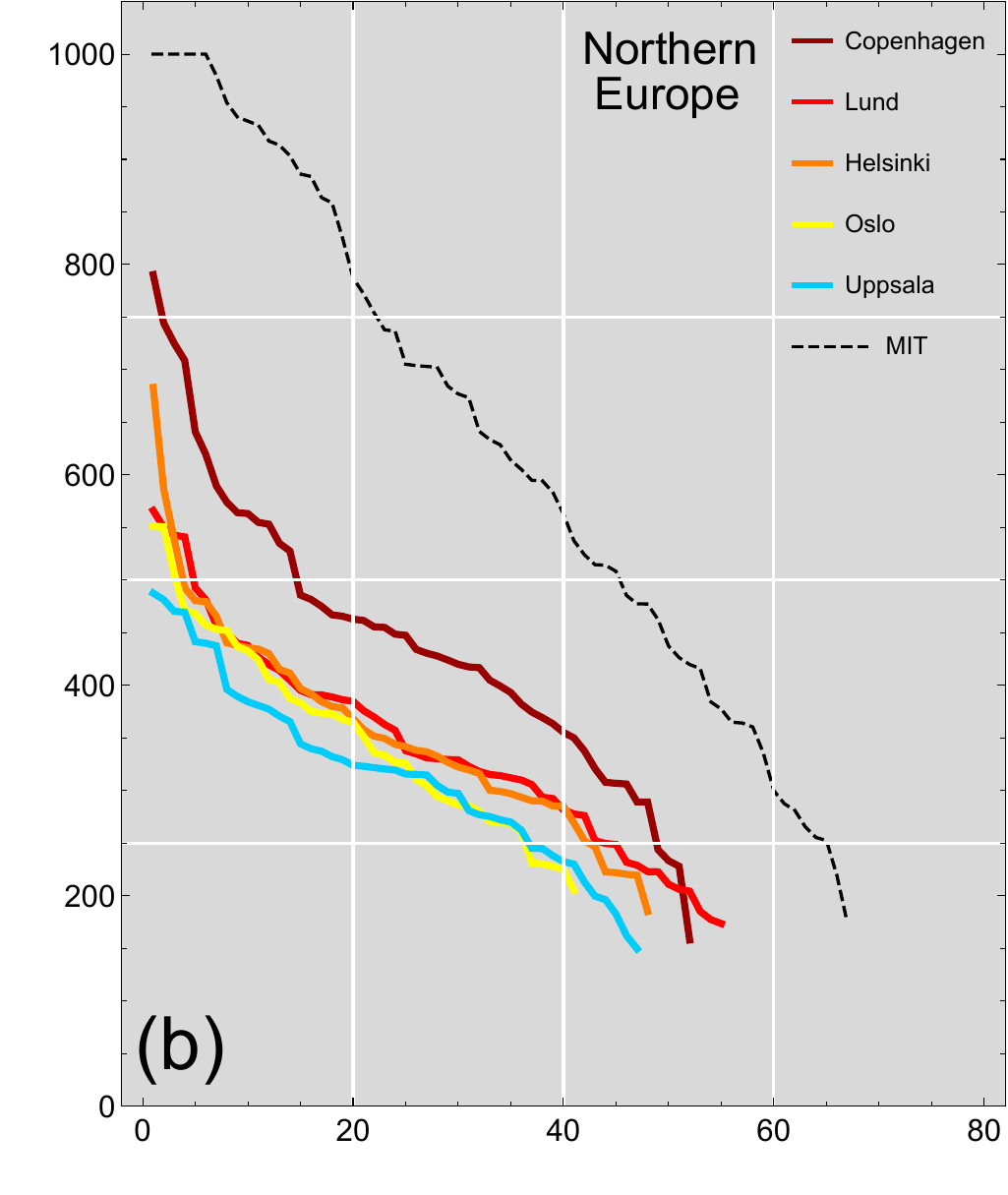}
\end{subfigure}
\begin{subfigure}[t]{0.33\textwidth}
    \includegraphics[scale=0.3]{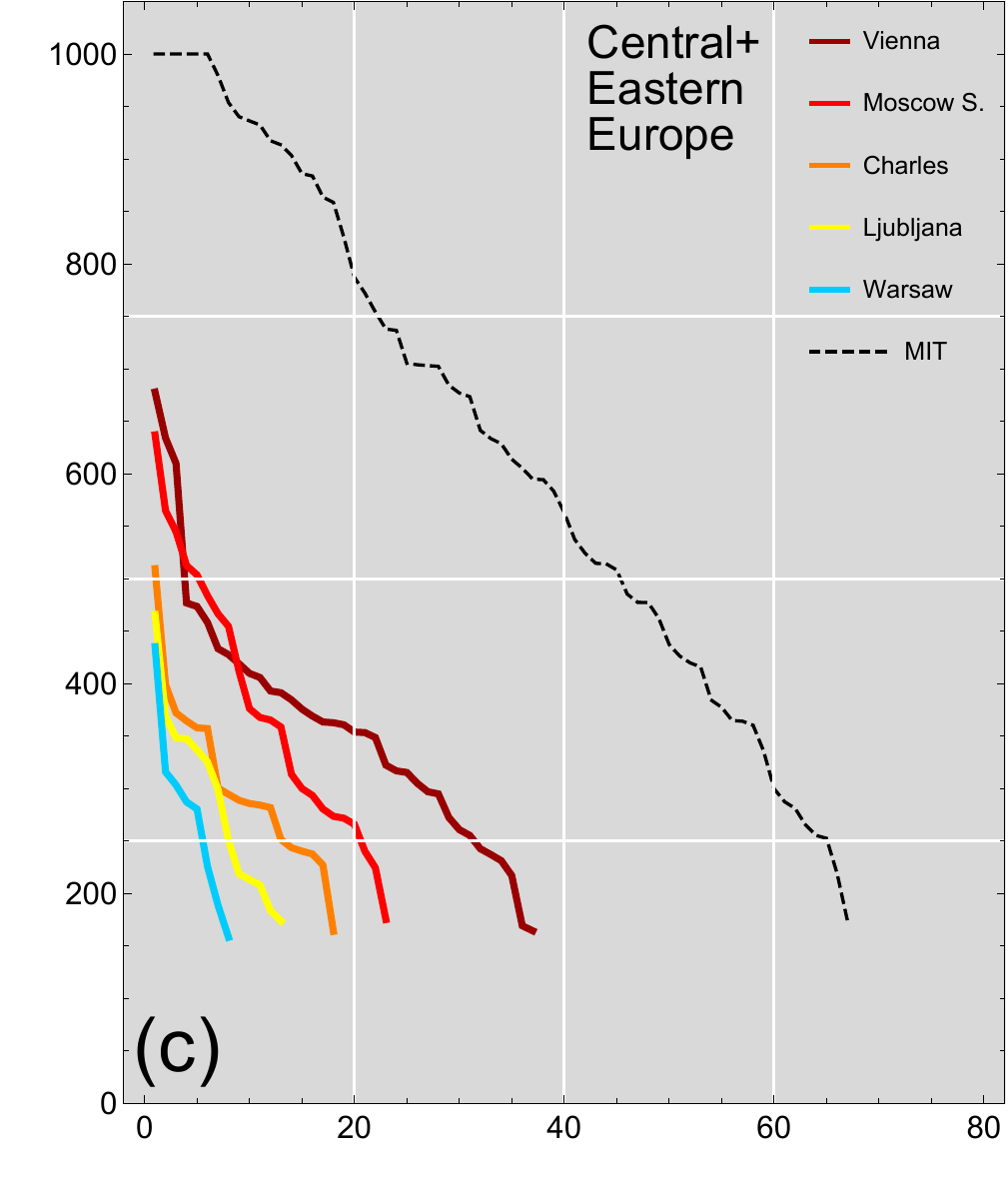}
\end{subfigure}

\begin{subfigure}[t]{0.33\textwidth}
    \includegraphics[scale=0.3]{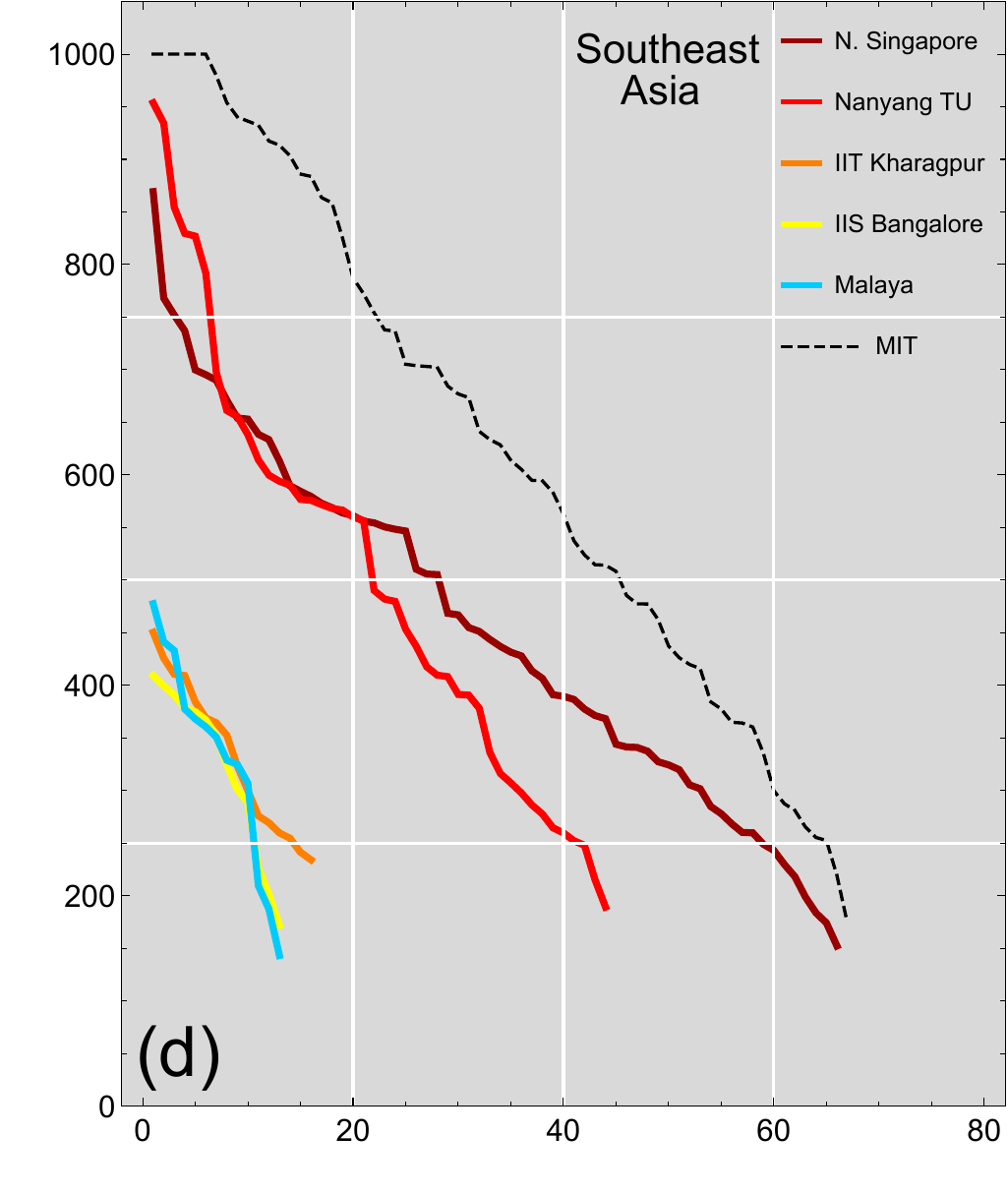}
\end{subfigure}
\begin{subfigure}[t]{0.33\textwidth}
    \includegraphics[scale=0.29]{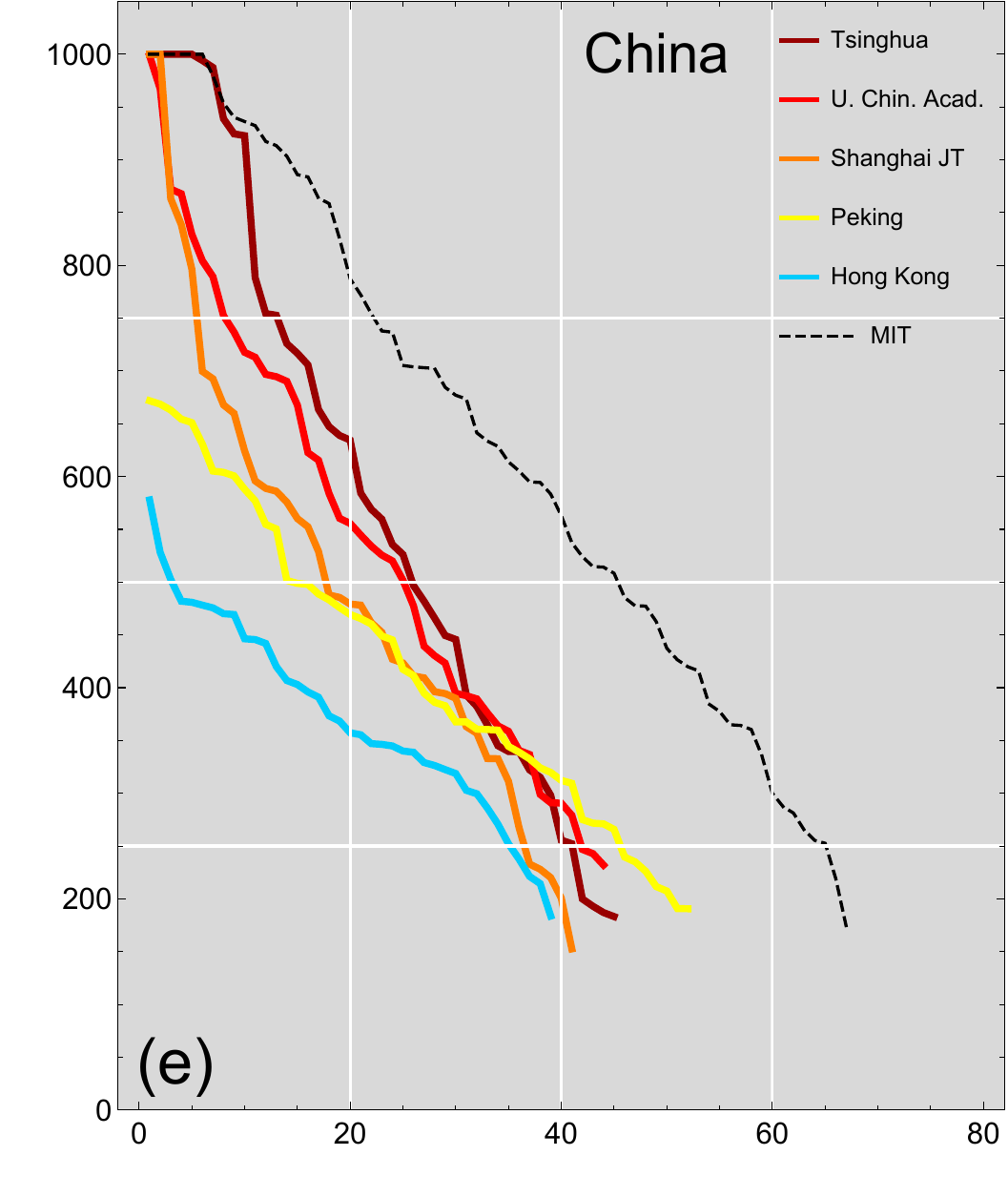}
\end{subfigure}
\begin{subfigure}[t]{0.33\textwidth}
    \includegraphics[scale=0.3]{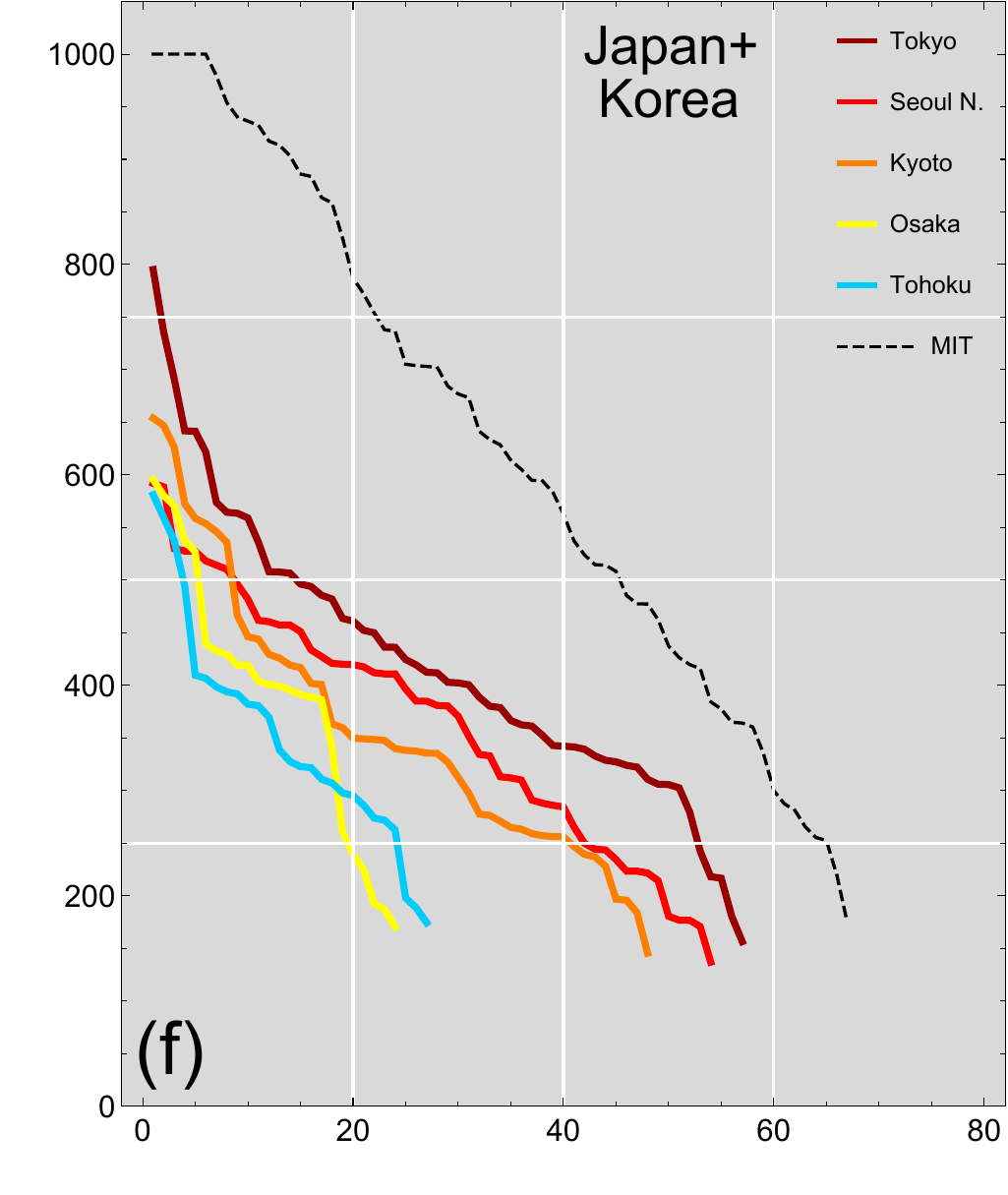}
\end{subfigure}

\begin{subfigure}[t]{0.33\textwidth}
    \includegraphics[scale=0.31]{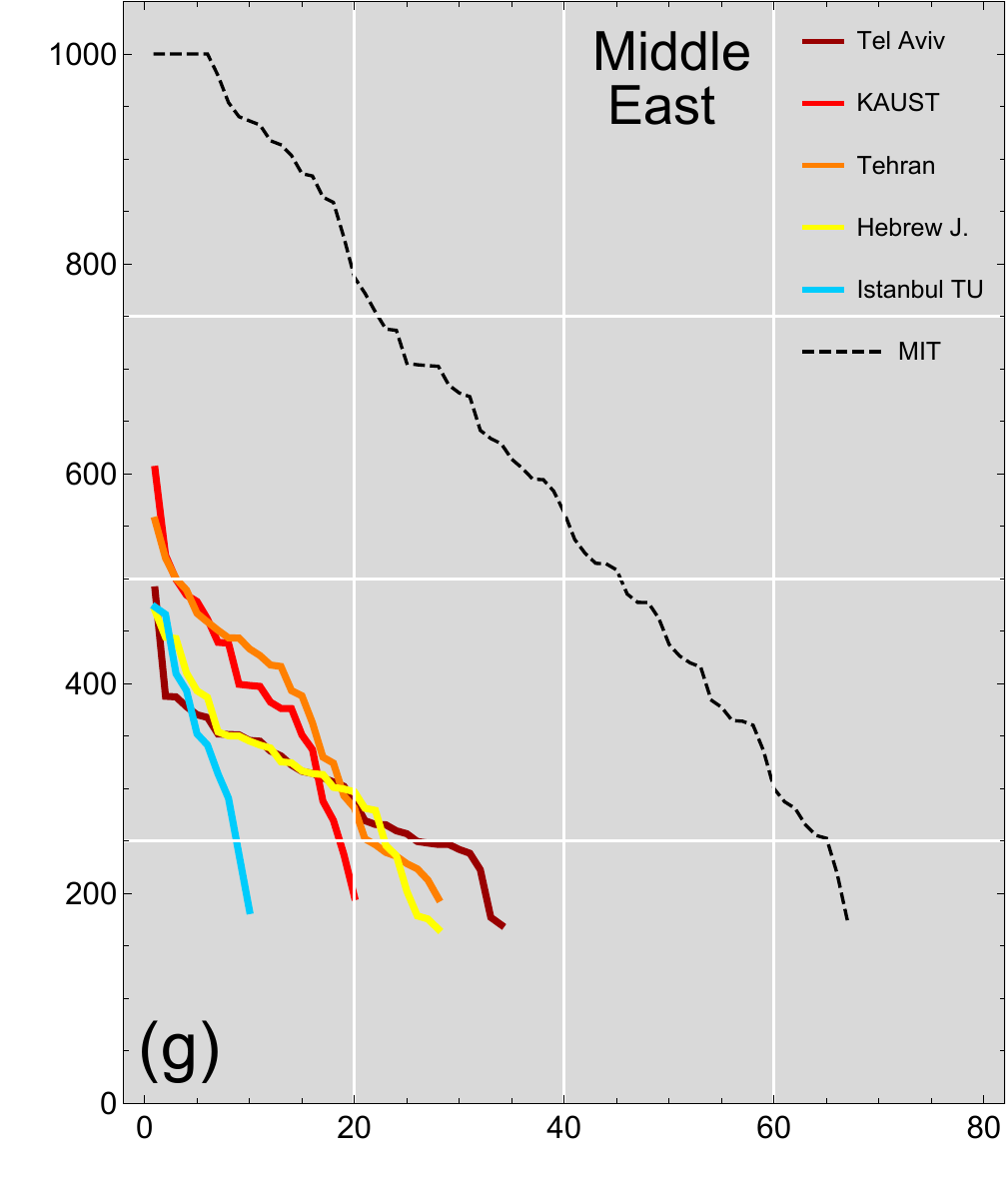}
\end{subfigure}
\begin{subfigure}[t]{0.33\textwidth}
    \includegraphics[scale=0.295]{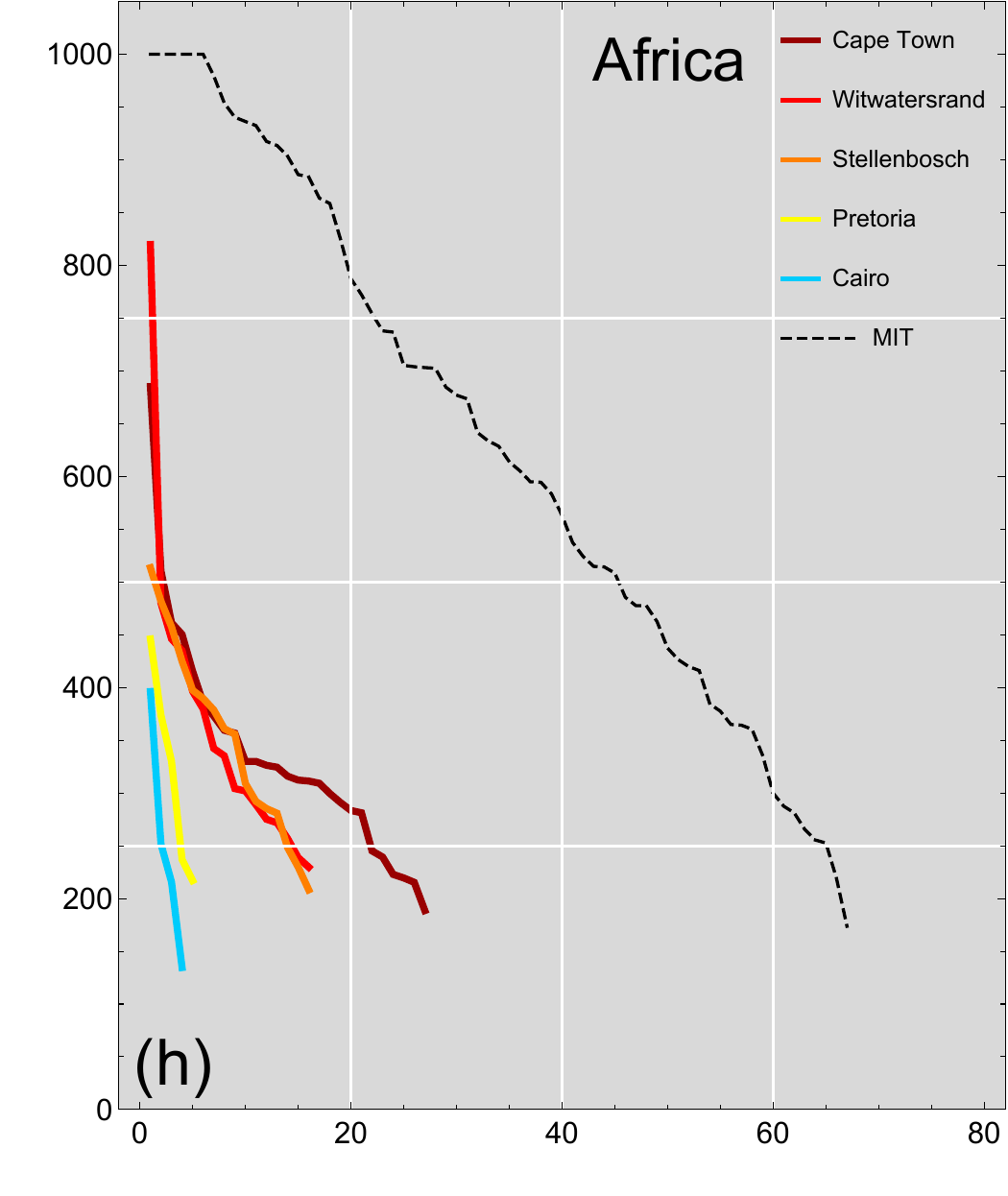}
\end{subfigure}
\begin{subfigure}[t]{0.33\textwidth}
    \includegraphics[scale=0.295]{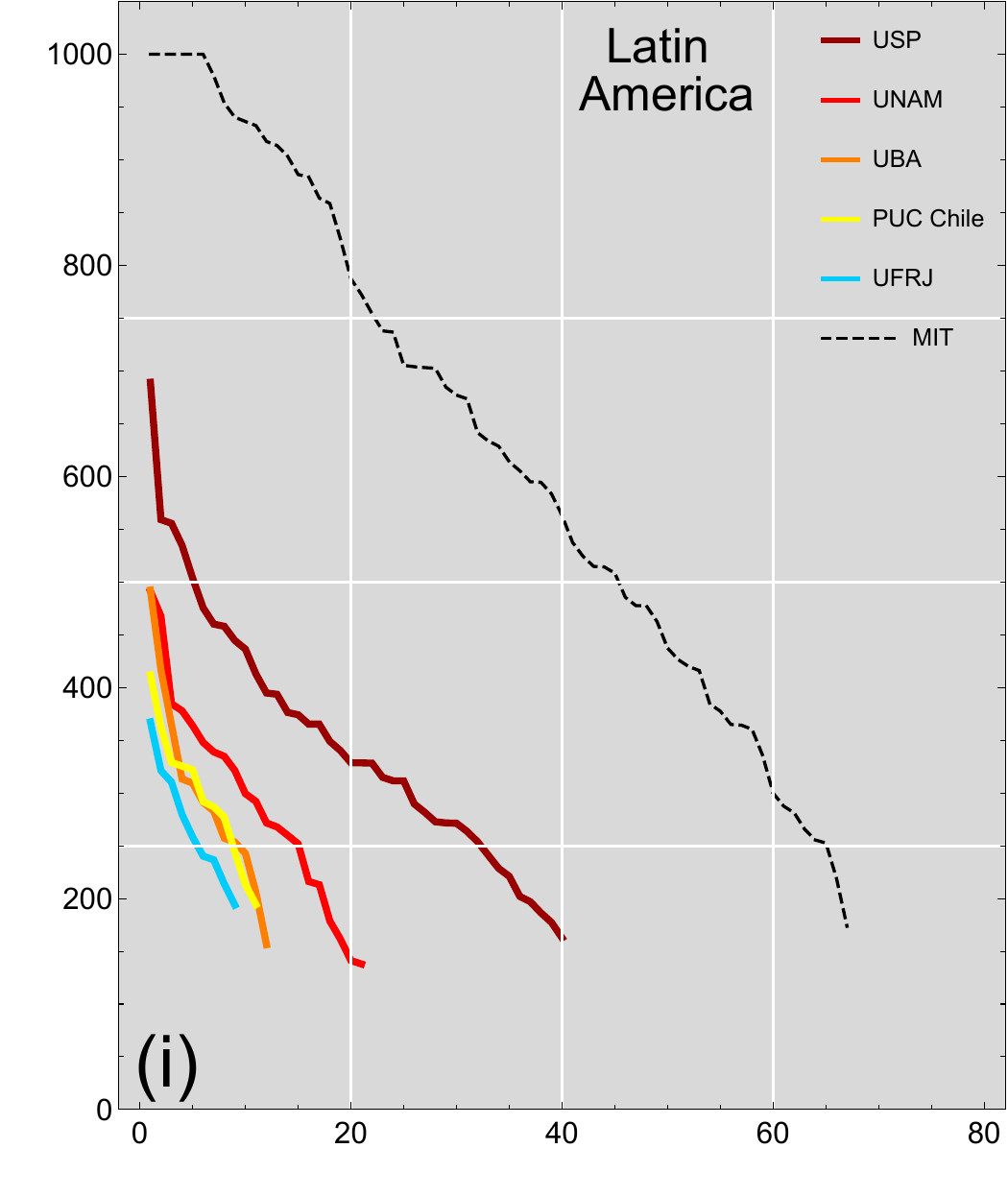}
\end{subfigure}
\caption{\vvg{Distribution of subject score (as in \jfm{Figure} \ref{fig:down1}) of major institutions of less developed regions in regards to research impact ordered from highest to lowest}.}
\label{fig:down2}
\end{figure}

\section{Regional \& Institutional Performance}

\y{Another relevant} aspect of the present analysis regards the regional and individual performance. In the case of comprehensiveness, \jfm{Figures} \ref{fig:down1}-\ref{fig:down2} depict the score of each institution (vertical axis) \y{of leading world} region\y{s} for each of the eighty subjects (horizontal axis). The higher $\mathcal{N}_{200}$ the farther the curve goes, and the higher classification in each subject ranking the higher curve in each plot. For instance, the upper bound of the overall score would entail a constant curve $\Lambda_{i}=1000$ in all subjects. \jfm{Figure} \ref{fig:down1}\jfm{a} shows that the best ranked institution, \textit{Harvard}, is still far from this upper bound. \y{In addition,} the similarity between all USA institutions belonging to group A++ is evident. Strikingly, the mean curve of this group seems to be well represented by the \textit{MIT} one. As such, in all remaining panels of \jfm{Figures} \ref{fig:down1}-\ref{fig:down2} I compare institutional performance against \y{that of} \textit{MIT}. \jfm{Figure} \ref{fig:down1}\jfm{b} demonstrates that many private USA institutions believed to be a peer of \textit{MIT} are indeed at a lower level (A+). \textit{Caltech}'s subject performance is significantly lower than its peers, but this is compensated by a much higher score per faculty. Conversely, \jfm{Figure} \ref{fig:down1}\jfm{c} shows \y{the leading} public USA institutions perform better than expected of their reputation, being at the same level of the private institutions of \jfm{Figure} \ref{fig:down1}\jfm{b}. In the United Kingdom, \jfm{Figure} \ref{fig:down1}\jfm{d} depicts the indiscernible performance of \textit{Oxford} and \textit{Cambridge} at par with \textit{MIT}, both belonging to the A++ group. Moreover, \textit{Imperial College} and \textit{UCL} are not too far behind. For the remaining panels (\jfm{Figures} \ref{fig:down1}\jfm{e-i} and \ref{fig:down2}), no institution will be found to be performing similarly to \textit{MIT} except for \textit{ETH Zurich} in Switzerland and the \textit{University of Toronto} in Canada. Among the leading countries in academic research, Germany seems to be underperforming in the subject rankings (see \jfm{Figure} \ref{fig:down1}\jfm{f}), but like \textit{Caltech} this is compensated by a much higher score per faculty density \y{than its peers}. In other words, German institutions perform a bit lower than its peers in leading European countries despite having a much smaller number of faculty per institution. For the remaining major regions of the world, a much lower performance is observed, except for the leading institutions in Denmark, Singapore, China, \y{and} Japan. \rx{The institutional performance amounts to the likelihood of their perfomance to be among the very best in any randomly chosen academic field. As shown in \jfm{figures} \ref{fig:Boxplot} and \ref{fig:qualitycomp}, very few similarity groups have a typical size-independent likelihood exceeding 40\%. It is the size of research faculty that ultimately strengthen a stratification of likelihood in performing well in any randomly chosen subject.}

\begin{figure}
\hspace{0.5cm}
\begin{subfigure}[t]{0.4\textwidth}
    \includegraphics[scale=0.4]{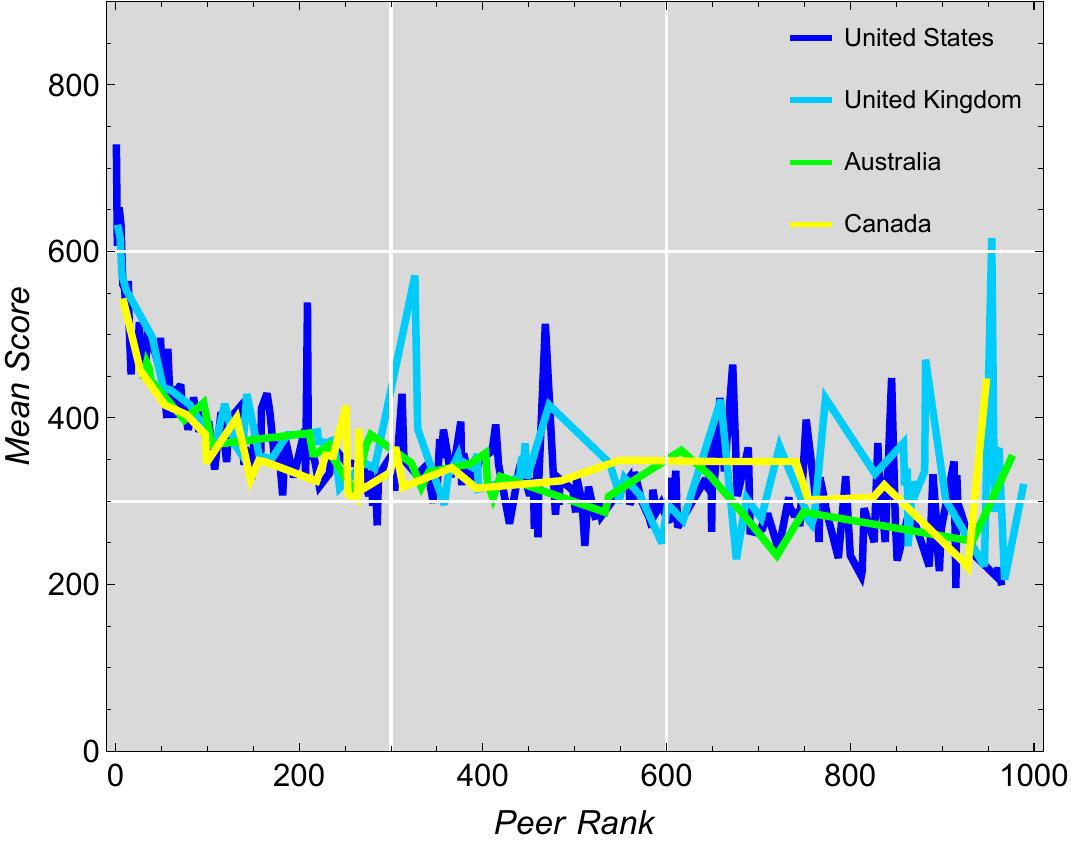}
\end{subfigure}
\hfill
\begin{subfigure}[t]{0.46\textwidth}
    \includegraphics[scale=0.4]{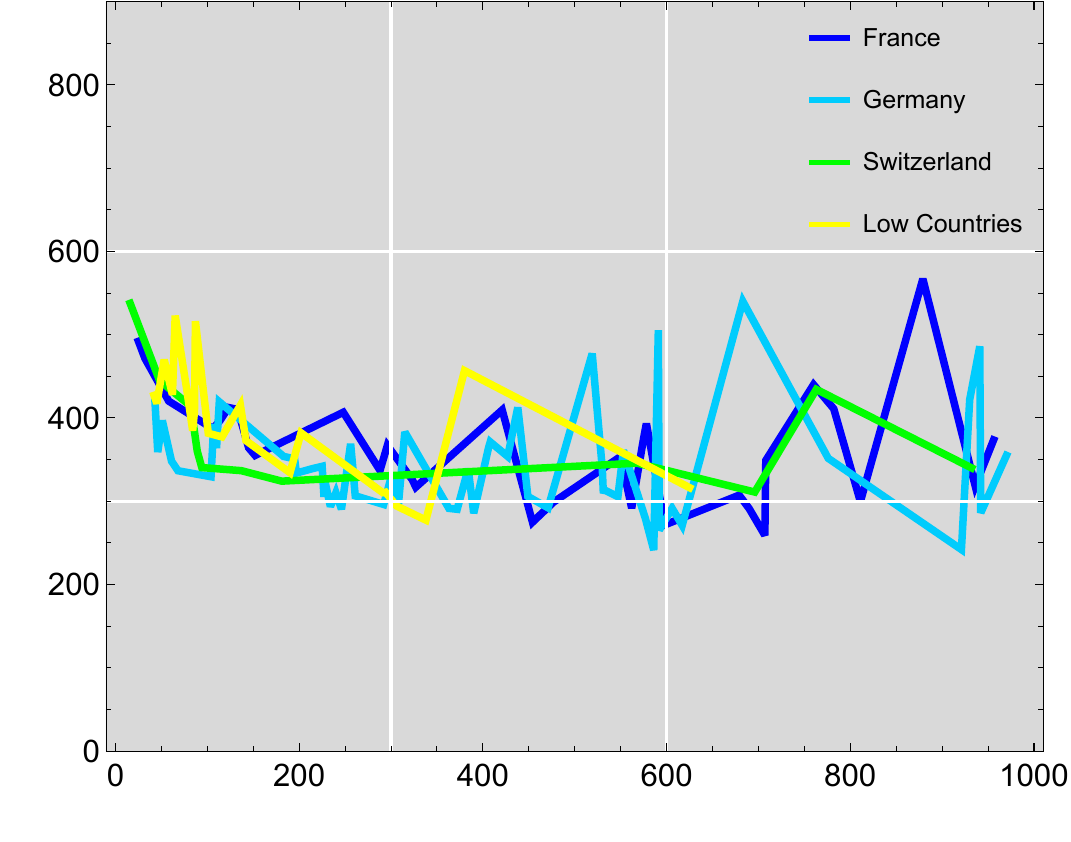}
\end{subfigure}

\hspace{0.5cm}
\begin{subfigure}[t]{0.4\textwidth}
    \includegraphics[scale=0.4]{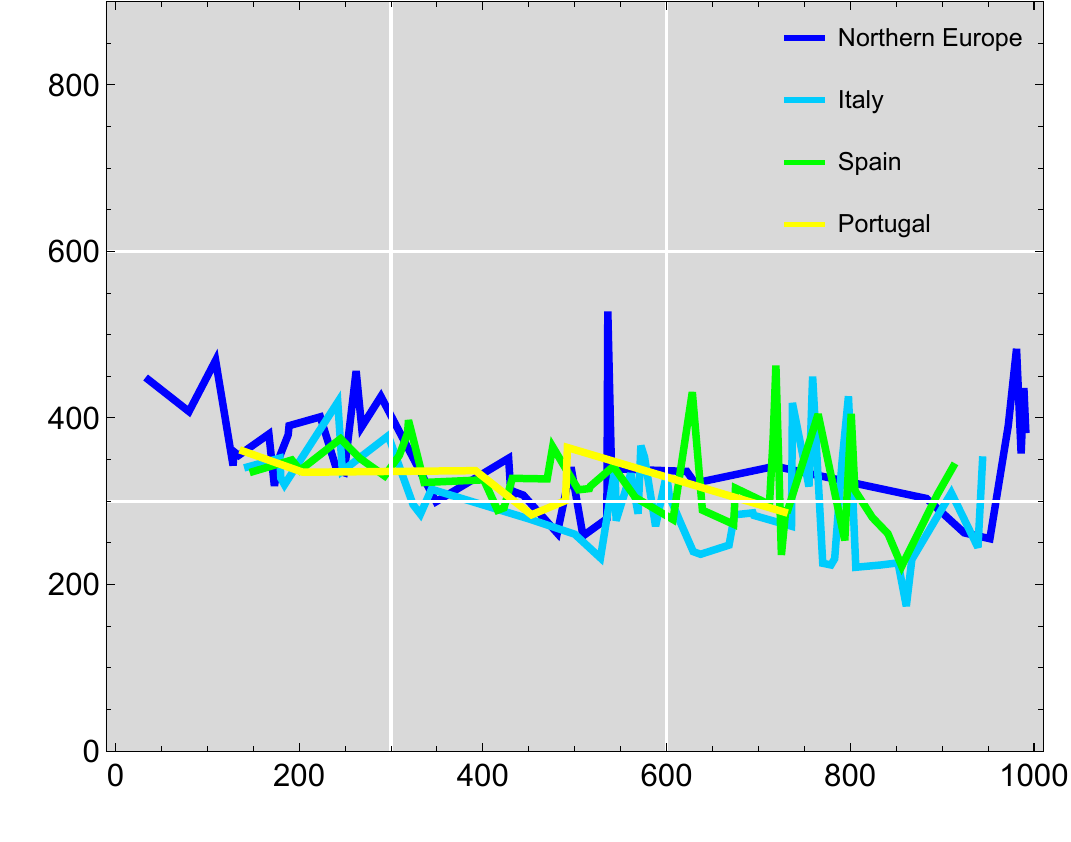}
\end{subfigure}
\hfill
\begin{subfigure}[t]{0.46\textwidth}
    \includegraphics[scale=0.4]{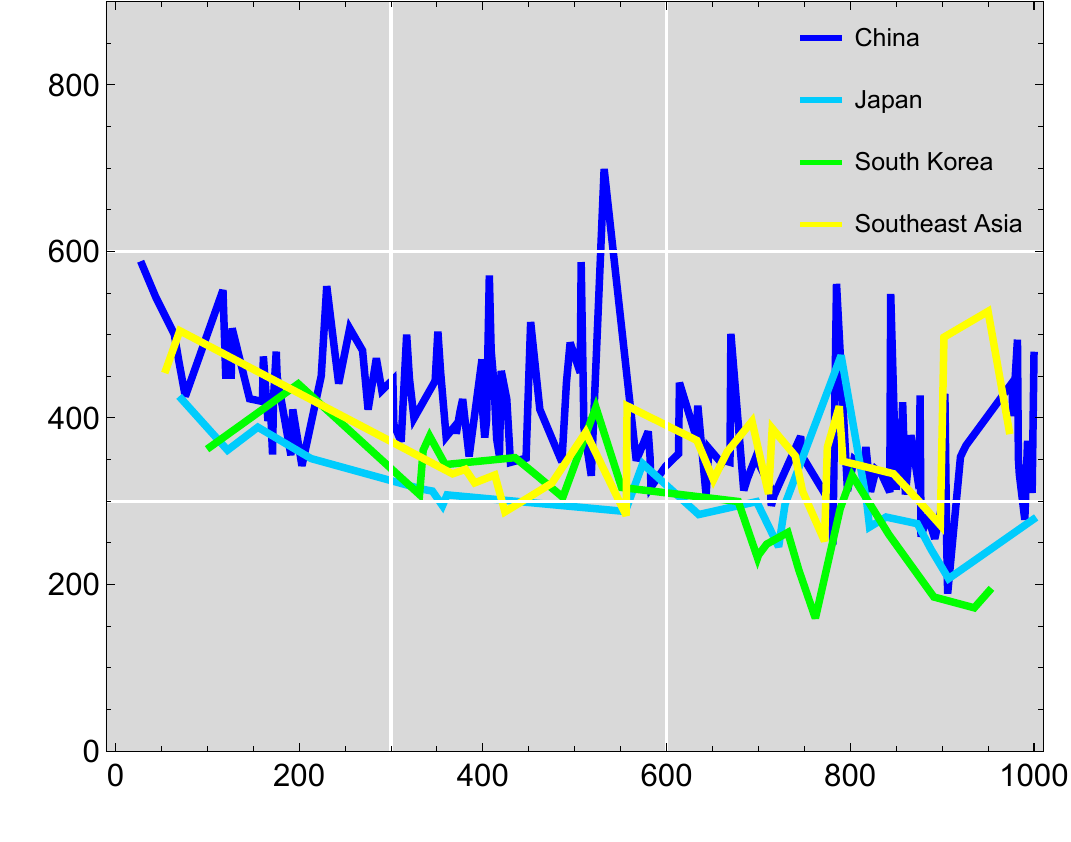}
\end{subfigure}

\hspace{0.5cm}
\begin{subfigure}[t]{0.4\textwidth}
    \includegraphics[scale=0.4]{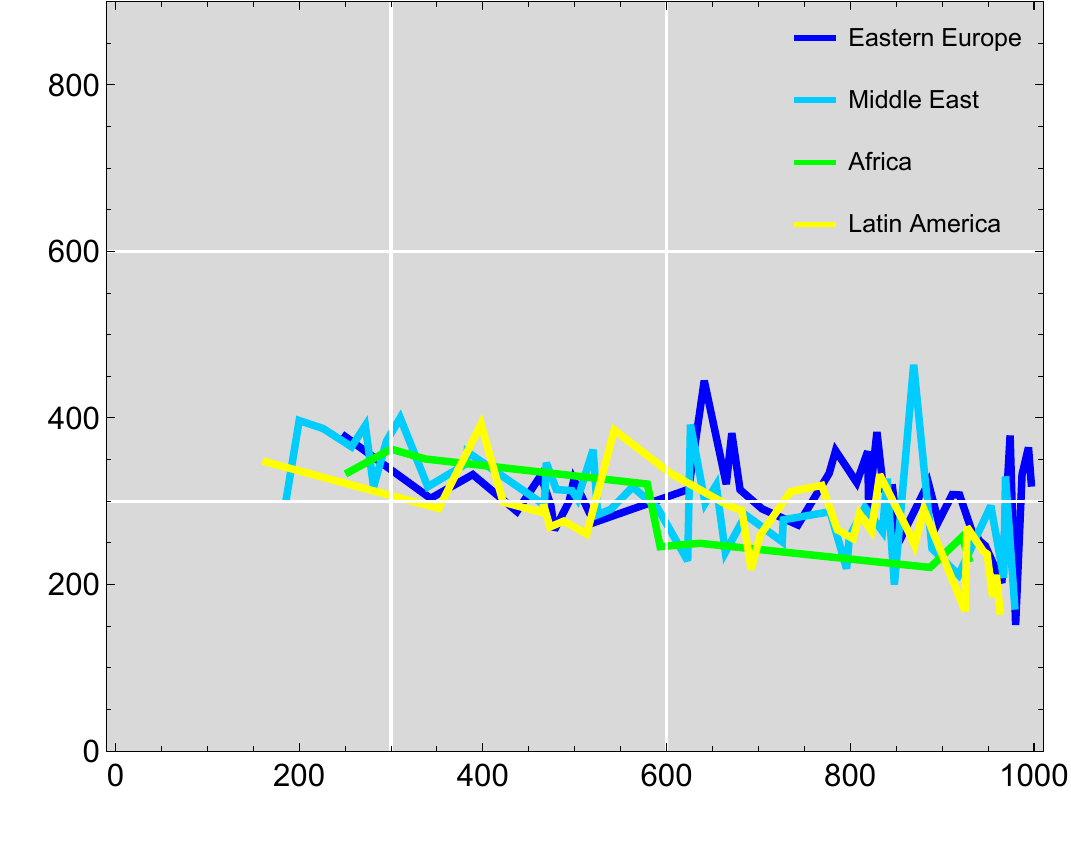}
\end{subfigure}
\hfill
\begin{subfigure}[t]{0.46\textwidth}
    \includegraphics[scale=0.36]{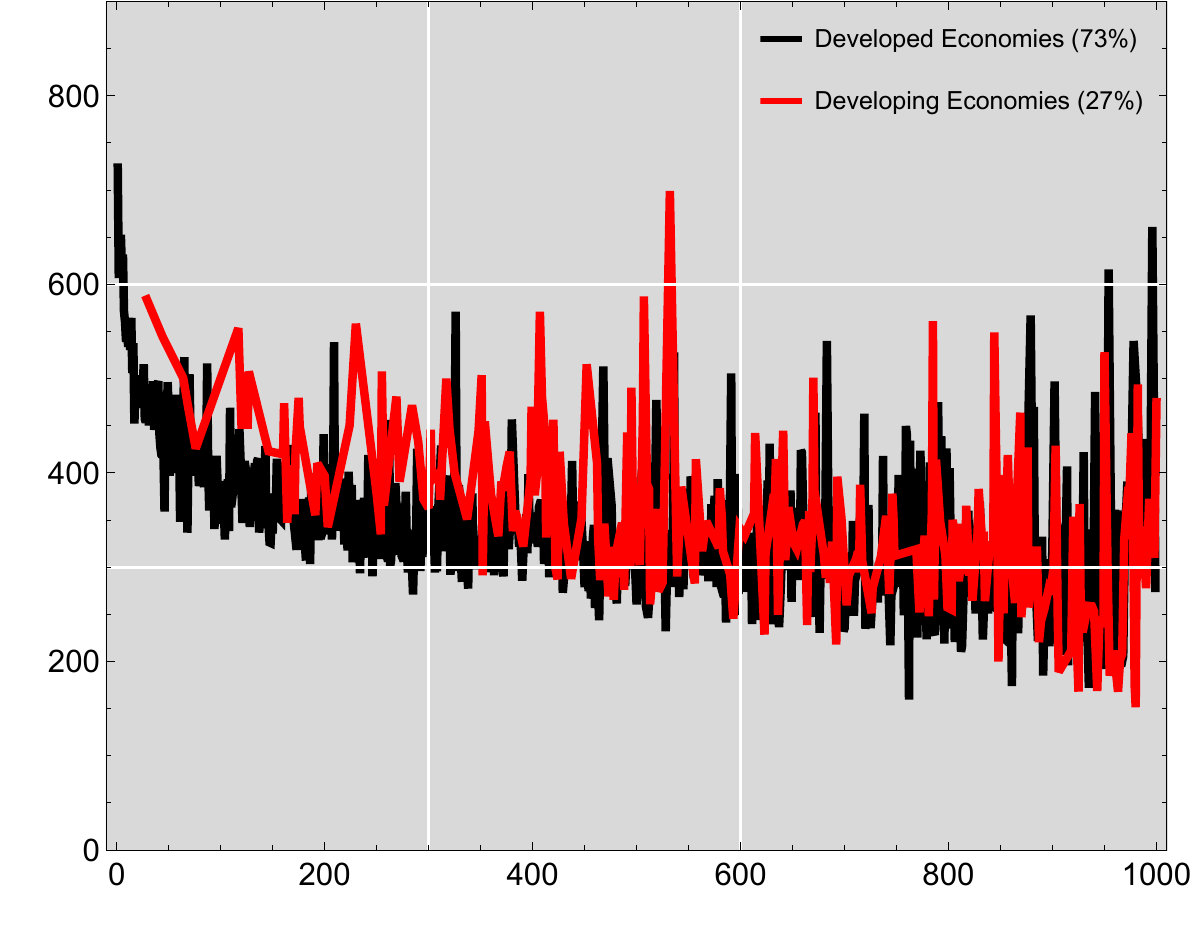}
\end{subfigure}
\caption{\vvg{Mean score across all academic subjects as a function of the peer rank for a variety of countries and regions. The last panel shows the break-down between developed and developing economies and the percentage of the total number of institutions \y{belonging to} each group.}}
\label{fig:regional}
\end{figure}

In \jfm{Figure} \ref{fig:regional} the regional performance of the mean score across all academic subjects is compared. The overall trend of saturation of mean score for $\mathcal{R}_{p} \gtrsim 200$ is found in all regions of the globe and their differences are negligible. Therefore, I conclude that not only quality is nearly homogeneously distributed across the leading thousand institutions, \y{it is also} regionally or \y{even nationally}. \y{Surprisingly}, academic quality is found to be worldwide homogeneous even despite economic disparities\footnote{The definitions for developed and developing economies can be found at the \href{https://www.imf.org/external/pubs/ft/weo/2022/01/weodata/groups.htm}{IMF report}.}, and the only difference between countries/regions is the total number \y{of} their institutions \y{and comprehensiveness thereof}.

\section{Conclusions}

The academic world saw in the past two decades an explosion of university league tables and their impact. The question regarding the liability for the assessment of university league tables arose. As \citet{weingart2005} argues, it is the duty and responsibility of peer-review alone to evaluate institutions and individuals in the research realm. However, as far as university tables are concerned, scientists rarely offer solutions or alternatives to the established frameworks. \vvg{Addressing such summons, I have attempted to answer the most pressing questions \y{on biases and negative features of league tables, overcoming them. Tangible} features of the institutions deemed to be "world-class" and the relationship of their metrics to the remaining research-intensive institutions were found.}

To the author's knowledge, the present work on university ranking is the first to \vvg{demonstrate that} status \y{development} \vvg{without concurrent\y{ly} affecting the position of} peer \vvg{institutions} is possible. Strong evidence supporting the stratification of excellence \y{in} academic performance has been found. \vvg{\y{Moreover, w}e solve the problem of the} \textit{zero-sum game} \vvg{through a} parameteriz\vvg{ation of the} research quality and output by implementing the idea of how broad the institution coverage of all main subjects of human knowledge are. Although most rankings are based on relative or comparative quality, the present methodology clearly classifies excellence by absolute values. The heart of the present rationale lies in the fact that normalized for subject average h-index, \y{whether} the research output \y{or any similar measure} of the leading institution in \y{a given subject} has the same influence and impact \y{of} the leading institution in \y{another subject}. \vvg{Thus, the} methodology \y{points to} the existence of groups of universities \y{in which performance} within each group \y{is almost} indistinguishable. Universities in the same similarity group can be considered equal and stratification within groups are hardly found.

\vvg{Albeit the stratification of academic excellence is real \y{between groups}, the present results} discredit the typical league table view of scarcity of reputation. The \y{analysis carried out in this study} has shown that nearly every \vvg{research-intensive} university excel\vvg{s} in a wide range of subjects. \vvg{Furthermore}, it has been unveiled that reputation and prestige often attached to awards and prizes in these fields, is not a proxy for quality. Unsurprisingly, I find that some of the \textit{Ivies} and similar private institutions in the United States do not measure up to their reputation and prestige. The discrepancy between actual performance and reputation is particularly substantial for \textit{Princeton} and \textit{Caltech}. On the other hand, the opposite seems to be true for large public institutions in North America, as their reputation does not measure up to their excellent performance and broad coverage \y{such as the \textit{University of Michigan} and the \textit{University of Wisconsin} flagship campuses}. Remarkably, \vvg{the strong correlation between the \textit{peer ranking} and comprehensiveness of universities is somehow also applicable to five established league tables. As such, I conclude that regardless of the methodology, research-intensive institutions tend to be classified as a function of their comprehensiveness.}

\vvg{The results of the present analysis may be perceived as a paradox: On one hand, the quality of an institution is not correlated with comprehensiveness. On the other hand, the\vvg{re is a} strong correlation between \vvg{comprehensiveness and the} mean ranking of the major established league tables. Nevertheless, it should by no means induce the reader to conclude that comprehensiveness \vvg{is necessary} to \vvg{make a good} university. \vvg{Rather, university league tables will inherently and even unconciously favor comprehensive institutions through their several differing methods, while the mean quality of all these institutions are nearly homogeneous. Hence, comprehensiveness is not a goal to be sought, but the root in the stratification of academic impact and unrelated to academic quality. In fact, comprehensiveness is a proxy for privilege, because the former requires large and steady endowment over decades in order to be achieved. Century-old institutions have a clear advantage in achieving a high level of comprehensiveness,
in particular those in countries with generous science funding bodies.}}

\section*{Acknowledgements}

The author thanks Jérôme Kasparian for reading the manuscript and adding to its readability.

\section*{Conflict of interest}

The author declares no conflict of interest.

\section*{Data Availability}

Core data can be found within personal \href{https://drive.google.com/file/d/1NJtWkS_vuLJg7pojutMGH3_W2Tdy3cSs/view?usp=sharing}{storage}. This includes links to search queries for each of the eighty subjects, league tables of the overall and subject rankings and Google Earth interactive map with 800 universities in the same color stratification within the manuscript.

%
%
%



\bibliographystyle{elsarticle-harv}
\bibliography{template}

\end{document}